\documentclass[final,3p,times]{elsarticle}
\makeatletter
\def\ps@pprintTitle{%
 \let\@oddhead\@empty
 \let\@evenhead\@empty
 \def\@oddfoot{}%
 \let\@evenfoot\@oddfoot}
\makeatother

\usepackage{amssymb}
\usepackage{float}
\usepackage{amsthm}
\usepackage{amsmath}
\usepackage{bm}
\usepackage{comment}
\usepackage{xcolor}
\usepackage{tabularx,ragged2e,booktabs,caption}

\newcommand\gammar{{\gamma_{\mathrm{ref}}}}
\newcommand\gammap{{\gamma_{\mathrm{TPS}}}}

\newcommand\Tw{{T_{\mathrm{w}}}}

\newcommand\Twr{{T_{\mathrm{w}}^{\mathrm{ref}}}}
\newcommand\qwr{{q_{\mathrm{w}}^{\mathrm{ref}}}}
\newcommand\Twp{{T_{\mathrm{w}}^{\mathrm{TPS}}}}
\newcommand\qwp{{q_{\mathrm{w}}^{\mathrm{TPS}}}}

\newcommand\Ps{{P_{\mathrm{s}}}}
\newcommand\Pst{{P_{\mathrm{d}}}}
\newcommand\Hd{{H_\delta}}
\newcommand{\Esp}[1]{{\mathbb{E}\left[#1\right]}}
\newcommand{\Var}[1]{{\mathbb{V}\left[#1\right]}}

\journal{Applied Mathematical Modelling}

\begin{document}

\begin{frontmatter}



\title{A surrogate-based optimal likelihood function for the Bayesian calibration of catalytic recombination in atmospheric entry protection materials}

\author[label1,label2]{Anabel del Val\corref{cor1}} 
\author[label3]{Olivier Le Ma\^{i}tre}
\author[label1]{Thierry E. Magin}
\author[label1]{Olivier Chazot}
\author[label2]{Pietro M. Congedo}
\address[label1]{von Karman Institute for Fluid Dynamics, Chauss\'{e}e de Waterloo 72, 1640 Rhode-St-Gen\`{e}se, Belgium}
\address[label2]{INRIA, Centre de Math\'{e}matiques Appliqu\'{e}es, \'{E}cole Polytechnique, IPP, Route de Saclay, 91128 Palaiseau Cedex, France}
\address[label3]{CNRS, INRIA, Centre de Math\'{e}matiques Appliqu\'{e}es, \'{E}cole Polytechnique, IPP, Route de Saclay, 91128 Palaiseau Cedex, France}
\cortext[cor1]{Corresponding author: von Karman Institute for Fluid Dynamics, Chauss\'{e}e de Waterloo 72, 1640 Rhode-St-Gen\`{e}se, Belgium. \textit{Email address:} ana.isabel.del.val.benitez@vki.ac.be (A. del Val)}

\begin{abstract}
This work deals with the inference of catalytic recombination parameters from plasma wind tunnel experiments for reusable thermal protection materials. One of the critical factors affecting the performance of such materials is the contribution to the heat flux of the exothermic recombination reactions at the vehicle surface. The main objective of this work is to develop a dedicated Bayesian framework that allows us to compare uncertain measurements with model predictions which depend on the catalytic parameter values. Our framework accounts for uncertainties involved in the model definition and incorporates all measured variables with their respective uncertainties. The physical model used for the estimation consists of a 1D boundary layer solver along the stagnation line. The chemical production term included in the surface mass balance depends on the catalytic recombination efficiency. As not all the different quantities needed to simulate a reacting boundary layer can be measured or known (such as the flow enthalpy at the inlet boundary), we propose an optimization procedure built on the construction of the likelihood function to determine their most likely values based on the available experimental data. This procedure avoids the need to introduce any a priori estimates on the nuisance quantities, namely, the boundary layer edge enthalpy, wall temperatures, static and dynamic pressures, which would entail the use of very wide priors. Furthermore, we substitute the optimal likelihood of the experimental measurements with a surrogate model to make the inference procedure both faster and more robust. We show that the resulting Bayesian formulation yields meaningful and accurate posterior probability distributions of the catalytic parameters with a reduction of more than 20\% of the standard deviation with respect to previous works. We also study the implications of an extension of the experimental procedure on the improvement of the quality of the inference.
\end{abstract}


\end{frontmatter}


\section{Introduction}
\label{Intro}

Space travel, since its beginnings in Low Earth Orbit (LEO) to the exploration of our Solar System, has led to countless scientific advancements in what it is one of the most challenging undertakings of humankind. Venturing into Space requires large amounts of kinetic and potential energy to reach orbital and interplanetary velocities. All this energy is dissipated when a space vehicle enters dense planetary atmospheres \cite{Laub}. The bulk of this energy is exchanged during the entry phase by converting the kinetic energy of the vehicle into thermal energy in the surrounding atmosphere through the formation of a strong bow shock ahead of the vehicle \cite{Anderson}. The interaction between the dissociated gas and a reusable protection system is governed by the material behavior which acts as a catalyst for recombination reactions of the atomic species in the surrounding gas mixture \cite{Catalysis}. The determination of the catalytic properties of thermal protection materials is a complex task subjected to experimental and model uncertainties, and the design and performance of reusable atmospheric entry vehicles must account for these uncertain characterizations.

It is relatively common when dealing with complex physical phenomena to resort to simple, non-intrusive a priori forward uncertainty propagation techniques~\cite{Turchi1, Mariotti}. These techniques assume a priori probability distributions for the main model parameters. Sensitivity analyses are then performed to discriminate the important ones. They also assume that the exact value is sufficiently well known and within the considered uncertainty range. These methods do not use any experimental observation to calibrate such parameters. The interest of using experimental information is that it leads to objective uncertainty levels and provides likely values rather than a priori guesses, achieving better and more reliable predictions.

In the present work, we explore the possibility of exploiting experimental data resulting from measurements in \cite{Panerai} for the purpose of inferring surface recombination efficiencies. These parameters play an important role in the prediction of the thermal response of selected protection materials, such as ceramic matrix composites. The inference focuses on a Bayesian approach that has the advantage of providing a complete characterization of the parameters' uncertainty through their resulting posterior distribution. While conceptually simple, performing a Bayesian inference raises several computational and practical difficulties at every one of its constitutive steps \cite{Weiping, Emery}. The main issue in our problem is related to the appearance of nuisance parameters within the model, such as pressures, wall temperatures and the boundary layer edge enthalpy. These particular parameters are needed to perform the inference but we are not explicitly interested in getting their distributions, nor we can measure all of them. Traditional Bayesian approaches deal with this problem by prescribing prior distributions on such parameters at the expense of some of the observations consumed to evaluate these nuisance parameter posteriors. Consequently, it is important to remark their impact on the quality of the inference \cite{Wirgin}.

Particularly, knowing how the experimental procedure is carried out is fundamental for the efficient formulation of the inference method. Testing for the characterization of a given protection material requires the use of an additional material which assists in the testing (referred to as reference material) and whose catalytic properties are better known. The Thermal Protection System (TPS) material in question and the reference material are then tested under the same experimental conditions. The basis of this experimental approach lays in the fact that we can have accurate knowledge about the inlet boundary condition if we immerse a well-characterized material in it for which the response, that it is directly measured, is dependent on those inlet conditions. This knowledge is then used to assess the properties of the TPS material, main objective of the experiment.

The inversion problem for the catalytic properties was first proposed in \cite{Sanson}. In it, Sanson et al. point out that the parameter for the reference material is not always perfectly known, and the testing conditions are subjected to uncertainties. These conditions are consequently treated as nuisance parameters in the inference problem by prescribing their priors. We highlight difficulties faced in this past work and propose a new methodology. Our formulation involves a particular treatment of the nuisance parameters, whose uncertainty is reduced by solving an auxiliary maximum likelihood problem. This maximum likelihood problem alleviates the need to sample the nuisance parameters, and can then improve the computational efficiency of the inference, providing more consistent and accurate posterior distributions. Solving this auxiliary problem and sampling the posterior distribution is expensive, as it requires multiple evaluations of the boundary layer equations. To mitigate this issue, we use a surrogate model of the optimal likelihood function, making the whole inference process faster and allowing for extensive exploration of the posterior distribution. The use of this methodology, novel in the field of TPS characterization, leads to an improved exploitation of the experimental measurements with, as a result, a better estimation of the catalytic parameters for a wide range of conditions. Further, the developments proposed in this paper have the potential of quantitatively assessing several ways in which to improve the experimental procedure and achieve a better understanding of the catalytic phenomena.

The article is organized as follows. Section 2 describes the experimental set-up together with the model-based simulations. In Section 3, the Bayesian framework is presented in detail. Section 4 looks into a case study with real experimental data to assess the validity of the Bayesian approach and study the different surrogate models for the log-likelihood approximation. Section 5 extends the methodology to other experimental cases to assess what the Bayesian method can bring to the problem of testing catalytic materials. Finally, Section 6 summarizes the outcomes of the analyses and discusses the possible perspectives of the presented approach.

\section{Experimental set-up and theoretical models}\label{BLCode}

In this section, we briefly discuss the theoretical model and recall the experiments and measured quantities considered in this work. As a simple model of TPS material, we define the catalytic coefficient $\gamma$ as the ratio of the number of atoms that recombine on the material surface over the total number of atoms that hit it. We assume the same recombination probability for the nitrogen and oxygen species constituting the air plasma, leading to just one single catalytic parameter to characterize the material under atmospheric entry conditions.
Model-based numerical simulations include this parameter to account for catalytic effects in the prediction of relevant quantities. The estimation of $\gamma$ also requires performing experiments in conditions relevant and similar to the environment faced during atmospheric entry so that this information can be fed to the model-based simulations to provide boundary conditions and closure. In all this sophisticated machinery, experiments and models are intertwined in a complex fashion, and it is essential to carefully assess the effect of possible uncertainties on the inferred quantities. 

\subsection{Experimental set-up}
\label{experiments}

\begin{figure}[!hbt]
\centering
\includegraphics[width=0.6\textwidth]{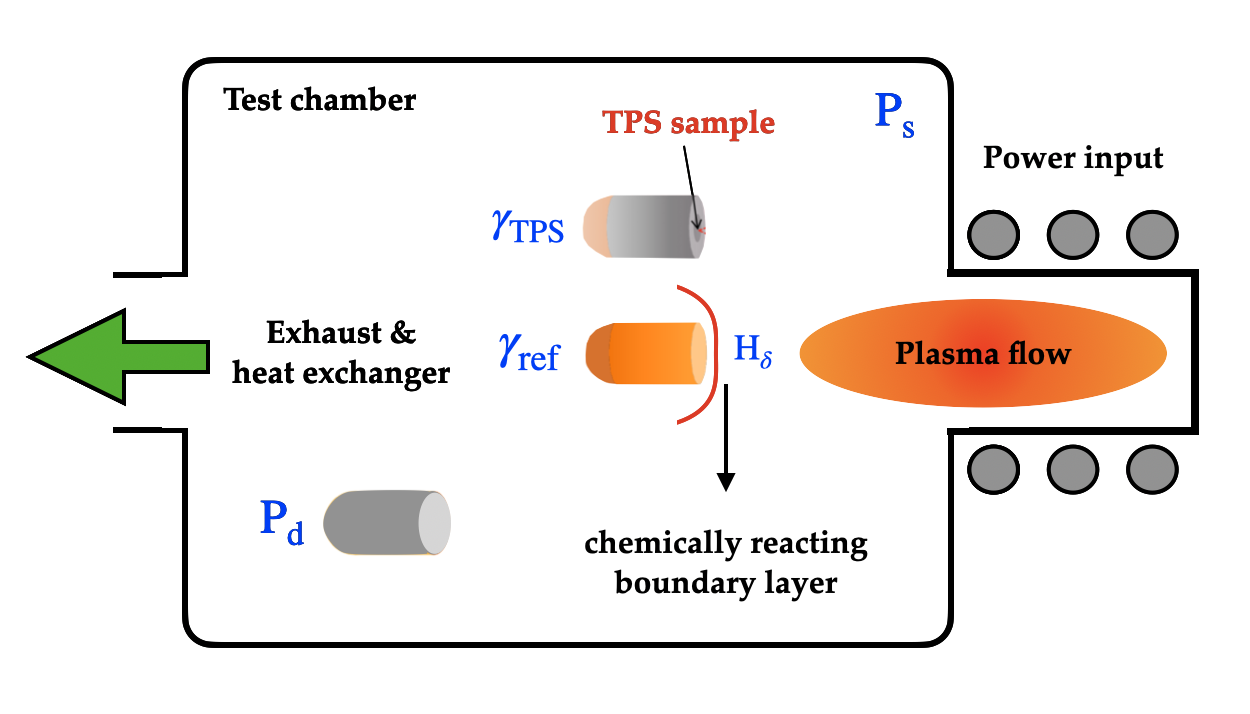}
\caption{Schematic view of the experimental set-up for the Plasmatron facility.}
\label{fig:3}
\end{figure}

We consider the experimental set-up of the Plasmatron facility at the von Karman Institute (VKI), a Inductively-Coupled Plasma (ICP) wind tunnel powered by a high-frequency, high-power, high-voltage (400 kHz, 1.2 MW, 2 kV) generator~\cite{Plasmatron}. 
Figure~\ref{fig:3} schematizes the Plasmatron and its instrumentation for catalytic property determination. 

The plasma flow is generated by the induction of electromagnetic currents within the testing gas in the plasma torch (right part of the scheme in Fig.~\ref{fig:3}); this process creates a high-purity plasma flow which leaves the testing chamber through the exhaust (left side of the scheme). In a typical experiment, one sequentially exposes two probes to the plasma flow: a reference probe made of a well-known material (copper), having a catalytic coefficient $\gammar$, and a test probe which holds a sample of the TPS material with the unknown catalytic coefficient, $\gammap$, to be inferred. The following instruments equip the Plasmatron. For pressures, a water-cooled Pitot probe measures the dynamic pressure $\Pst$ within the plasma jet, and an absolute pressure transducer records the static pressure $\Ps$ in the Plasmatron chamber. The reference probe is an hemispherical device (25 mm radius) equipped with a water-cooled copper calorimeter at the center of its front face. The calorimeter has a cooling water system that maintains the surface temperature of the reference probe $\Twr$. The heat flux $\qwr$ is deduced from the mass flow (controlled by a calibrated rotameter) circulating in the cooling system and the inlet/outlet water temperature difference measured by thermocouples as a result of the exposure to the plasma flow. For the test probe, we measure directly the heat flux $\qwp$ and surface temperature $\Twp$. The determination of the heat flux  assumes a radiative equilibrium at the surface, with the relation $\qwp =\sigma \epsilon \Twp^{4}$, where $\sigma$ is the Stefan-Boltzmann constant, $\epsilon$ is the emissivity measured with an infrared radiometer, and $\Twp$ is the wall temperature which is measured using a pyrometer. More details on how these measuring devices work can be found in~\cite{Helber}. 

The underlying idea of the experimental procedure is to perform first measurements of the wall temperature $\Twr$, heat flux  $\qwr$ and pressures $\Pst$ and $\Ps$ with the reference probe set in the plasma jet. As these measurements depend on the state of the free stream flow, in particular on the enthalpy $\Hd$ at the boundary layer edge, the free stream conditions can be deduced if one knows the catalytic coefficient $\gammar$ of the reference probe. Then, in a second stage, the test probe is set in place of the reference probe in the plasma jet. The corresponding steady state wall temperature $\Twp$ and heat flux $\qwp$ are measured and, assuming that the free stream flow conditions have not changed, the catalytic coefficient $\gammap$ of the test probe can be inferred.

\subsection{Model-based simulations: the Boundary Layer code}\label{BL:code}
To identify the TPS catalytic properties $\gammap$, we simulate the chemically reacting boundary layer in the vicinity of the probe stagnation point~\cite{Anderson}. The Boundary Layer (BL) code solves the full Navier-Stokes equations along the stagnation line. To solve the system, we need closure models for the thermodynamic and transport properties as well as the chemical production terms of the different species. Transport fluxes are derived from kinetic theory using the Chapman-Enskog method for the solution of the Boltzmann equation~\cite{Ferziger, Mitchner}. Diffusion fluxes are computed through the generalized Stefan-Maxwell equations, an approach which by derivation is exactly equivalent to a description based on the multicomponent diffusion matrix~\cite{Chapman, Giovangigli, Kolesnikov}, therefore providing equivalent solutions for the diffusion fluxes. For the homogeneous chemistry in the gas phase, the Law of Mass Action is used to compute production rates as proportional to the product of the reactant densities raised to their stoichiometric coefficients~\cite{Kenneth}. A 7-species air mixture and chemical rates from Dunn and Kang~\cite{Dunn} in the form of modified Arrhenius laws are considered. The thermodynamic properties, such as the enthalpy, are derived from statistical mechanics~\cite{Anderson, Vicenti} for a reacting mixture of perfect gases, assuming thermal equilibrium and chemical non-equilibrium. Apart from the closure models, the parabolic nature of the BL equations requires the imposition of two boundary conditions: the external flow conditions at the boundary layer edge, namely pressure, temperature and velocity, and the conditions at the material surface. A mass balance is imposed to account for the production and depletion of species as a consequence of their interactions with the material surface. Recombination reactions can be triggered depending on the catalytic activity of such material~\cite{Goulard}. We impose a no slip condition at the wall for the momentum equation and impose a wall temperature for the energy flux. More details about the derivation, coordinate transformations and numerical implementation of the BL code are available in the work of Barbante~\cite{Barbante}. In summary, the predictive quantity of the code is the wall heat flux 

\begin{equation}
\label{eq:heatfluxI}
    q_{\mathrm{w}} = q_{\mathrm{w}}\left(\gamma, T_{\mathrm{w}}, P_{\delta}, H_{\delta}, \delta, \dfrac{\partial u_{\delta}}{\partial x}, v_{\delta}\dfrac{\partial}{\partial y} \left(\dfrac{\partial u_{\delta}}{\partial x}\right)\right),
\end{equation}
which depends on the free stream conditions (subscript $\delta$), the thickness of the boundary layer $\delta$, the catalytic parameter of the material $\gamma$ and the surface temperature $T_{\mathrm{{w}}}$. An auxiliary magnetohydrodynamic axisymmetric simulation assuming Local Thermodynamic Equilibrium (LTE) is performed to simulate the flow in the torch and chamber of the wind tunnel~\cite{lani}. Relaying on the knowledge of the operating conditions of the Plasmatron, such as electric power, injected mass flow, static pressure and probe geometry, this 2D simulation lets us compute non-dimensional parameters that define the momentum influx to the boundary layer (interested reader is directed to~\cite{DegrezNDP}). To derive the computation of the velocity derivative $\partial u_{\delta}/\partial x$ needed as input in Eq.~\ref{eq:heatfluxI}, we need also to incorporate the measurement of the dynamic pressure $\Pst$ along with the non-dimensional parameters. The prediction we are seeking to match the experimental data is now recast as

\begin{equation}
\label{eq:heatfluxII}
    q_{\mathrm{w}} = q_{\mathrm{w}}\left(\gamma, T_{\mathrm{w}}, P_{\delta}, H_{\delta},\Pst,\mathrm{\Pi_{1},\Pi_{2},\Pi_{3}}\right),
\end{equation}
where $\mathrm{\Pi_{1},\Pi_{2},\Pi_{3}}$ are the non-dimensional parameters and $P_{\delta}$ is taken as the chamber static pressure $P_{s}$. These parameters, together with the dynamic pressure $\Pst$, define the boundary layer thickness, velocity and derivatives at the edge. Even though the number of input parameters is larger in Eq.~\ref{eq:heatfluxII}, this formulation is more useful given that we have replaced three unknown and not easily measurable quantities by a quantity that can be directly measured ($\Pst$) and three parameters that do not depend on the local flow conditions and can be taken as known constants for each condition for the purpose of our Bayesian formulation \cite{Sanson1}.

The typical procedure to retrieve the catalytic parameter $\gamma$ is the following. In the first step, for given non-dimensional parameters $\mathrm{\Pi_{1},\Pi_{2},\Pi_{3}}$, measured wall temperature $\Twr$, static pressure $\Ps$, dynamic pressure $\Pst$ and reference catalytic parameter $\gammar$, the enthalpy $\Hd$ that matches the heat flux observed $\qwr$ is iteratively determined. When the enthalpy $\Hd$ is determined, the code can be used, in a second step, to find the value of the catalytic parameter $\gammap$ that yields the observed heat flux $\qwp$ for the measured wall temperature $\Twp$, pressures $\Ps$ and $\Pst$ and rebuilt enthalpy $\Hd$ obtained in the previous step.


\section{Bayesian framework}
\label{Bayes}
In this section we derive the Bayesian formulation of the inference problem in subsection~\ref{bayesian:formulation}. We then discuss in subsection~\ref{Surrogates} the construction of a surrogate model to approximate the likelihood function, and finally, briefly describe the procedure used to sample the posterior distribution in subsection~\ref{mcmc}.

\subsection{Bayesian formulation of the inference problem}
\label{bayesian:formulation}

The inference of the model parameters uses the Bayes formula which can be generically formulated as
\begin{equation} \label{eq:Bayes}
P(\bm Q \vert \bm M)=\dfrac{L (\bm M \vert \bm Q) \ P (\bm Q)}{P_{\bm M}(\bm M)}.
\end{equation}
In~\eqref{eq:Bayes}, we have denoted $\bm Q$ the vector of model parameters to be inferred, $\bm M$ the vector of measurements or observations, $P(\bm Q)$ the prior distribution of the parameters, $L (\bm M \vert \bm Q)$ the likelihood of the measurements, $P(\bm Q \vert \bm M)$ the posterior distribution of $\bm Q$, and $P_{\bm M}$ the evidence or marginal likelihood, that is, the probability that the measurements are obtained under the considered model. In our context, we have  $\bm M = (P_{\mathrm{s}}^{\mathrm{meas}}, P_{\mathrm{d}}^{\mathrm{meas}},
q_{\mathrm{w}}^{\mathrm{ref},\mathrm{meas}}, q_{\mathrm{w}}^{\mathrm{TPS},\mathrm{meas}},T_{\mathrm{w}}^{\mathrm{ref},\mathrm{meas}}, T_{\mathrm{w}}^{\mathrm{TPS},\mathrm{meas}})$. Classically, the likelihood measures the discrepancies between the measurements in $\bm M$ and the corresponding model predictions. The issue here is that the model predictions are not just functions of the catalytic coefficients $\bm \gamma =(\gammar, \gammap)$, but also depends on all the inputs of the BL code: the pressures $\Ps, \Pst$, wall temperatures $T_{\mathrm{w}}^{\mathrm{ref}}, T_{\mathrm{w}}^{\mathrm{TPS}}$ and boundary layer edge enthalpy $H_{\delta}$. The pressures and wall temperatures are measured in the experiment, but only with limited precision, while the enthalpy $H_{\delta}$ is simply not known. Consequently, there may be zero, or multiple, boundary layer edge conditions consistent with the measurements. Since the boundary layer edge conditions can not be completely characterized, the remaining uncertainty should be accounted for when inferring the test probe catalytic coefficients.

One possibility to handle this issue is to consider the whole set of uncertain quantities, not just the quantities of interest $\gammar$ and $\gammap$, but also the so-called nuisance parameters. In that case, we define the vector of model parameters
$\bm Q = (\gammar, \gammap, T_{\mathrm{w}}^{\mathrm{ref}}, T_{\mathrm{w}}^{\mathrm{TPS}},\Ps, \Pst, H_{\delta})$ in the inference problem. The introduction of the nuisance parameters induces several difficulties related to the necessity to specify their prior distributions, the increased dimensionality of the inference space, and the consumption of information for the inference of the nuisance parameters. This last issue is detrimental to learning the parameters of interest. In~\cite{Sanson} non-informative priors were used for all the nuisance parameters. This approach only approximates the posterior of $\bm Q$ including the nuisance parameters, and the influence of the (unknown) prior densities of these parameters is unclear. Not only that but the ability to effectively learn from these experimental data is lost. In the following, we derive an alternative formulation for the joint inference of the two catalytic coefficients $\bm \gamma = (\gammar,\gammap)$. Specifically, we consider the following Bayes formula
\begin{equation} \label{eq:14}
P(\bm \gamma \vert \bm M)=\dfrac{L (\bm M \vert \bm \gamma) \ P(\bm \gamma)}{P(\bm M)},
\end{equation}
as before, $L (\bm M \vert \bm \gamma)$ refers to the likelihood of the  measurements in $\bm M$. This formulation only depends on the two catalytic coefficients $(\gammar, \gammap)$ and not on the other nuisance parameters. As a result, only the prior $P(\bm \gamma)$ is needed.

\subsubsection{Likelihood function}

Our objective is to design a reduced likelihood function which does not involve any nuisance parameters. As stated before, the prediction of the heat fluxes involves not only the catalytic coefficients $\gammar, \gammap$, but also $\Ps, \Pst, T_{\mathrm{w}}^{\mathrm{ref}}, T_{\mathrm{w}}^{\mathrm{TPS}}$ and $H_{\delta}$. Assuming independent unbiased Gaussian measurement errors, with magnitude $\sigma$, the full likelihood of $\bm M$ with the nuisance parameters would read as
\begin{equation} \label{eq:15}
\begin{split}
L(\bm M \vert \bm Q) = & 
\exp \left[-\dfrac{\left( P_{\mathrm{s}}^{\mathrm{meas}}-\Ps \right)^{2}}
{2 \sigma_{\Ps}^{2}}\right] \times 
\exp\left[-\dfrac{\left( P_{\mathrm{d}}^{\mathrm{meas}}-\Pst \right) ^{2}}{2 \sigma_{\Pst}^{2}}\right] \times\\
&\times \prod_{i \in \left\{\mathrm{ref, TPS}\right\}} \ \exp \left[-\dfrac{\left( q_{\mathrm{w}}^{i,\mathrm{meas}}-q_{\mathrm{w}}^{i}(H_{\delta}, \Ps, \Pst, T_{\mathrm{w}}^{i}, \gamma_{i}) \right) ^{2}}{2 \sigma_{q_{\mathrm{w}}}^{2}}
-\dfrac{\left( T_{\mathrm{w}}^{i, \mathrm{meas}}-T_{\mathrm{w}}^{i} \right) ^{2}}{2 \sigma_{T_{\mathrm{w}}}^{2}}\right].
\end{split}
\end{equation}
In this likelihood, the dependencies on $\gammar$ and $\gammap$ are implicitly contained in the heat flux terms. To reduce the dependencies of the likelihood to just the parameter $\bm \gamma$, we propose to set the nuisance parameters $\bm Q \setminus \bm \gamma$ to the values that maximize the likelihood~\eqref{eq:15}. 
In the following, we denote $H_{\delta}^{\mathrm{opt}}, P_{\mathrm{s}}^{\mathrm{opt}}, P_{\mathrm{d}}^{\mathrm{opt}}, T_{\mathrm{w}}^{i, \mathrm{opt}}$, the maximizers of~\eqref{eq:15}. Note that these optima are functions of the catalytic coefficients. We shall also denote $q_{\mathrm{w}}^{i, \mathrm{opt}}(\bm \gamma)$ the corresponding model predictions of the heat fluxes for each probe. With these optimal values for the nuisance parameters, we define the optimal likelihood as
\begin{equation} \label{eq:17}
\begin{split}
L^{\mathrm{opt}}(\bm M \vert \bm \gamma) = & \exp \left[-\dfrac{\left( P_{\mathrm{s}}^{\mathrm{meas}}-P_{\mathrm{s}}^{\mathrm{opt}}(\bm \gamma)\right)^{2}}{2 \sigma_{P_{\mathrm{s}}}^{2}}\right] \ \exp \left[-\dfrac{\left( P_{\mathrm{d}}^{\mathrm{meas}}-P_{\mathrm{d}}^{\mathrm{opt}}(\bm \gamma) \right) ^{2}}{2 \sigma_{P_{\mathrm{d}}}^{2}}\right] \times \\
&\times \prod_{i \in \left\{\mathrm{ref, TPS}\right\}} \ \exp \left[-\dfrac{\left( q_{\mathrm{w}}^{i, \mathrm{meas}}-q_{\mathrm{w}}^{i, \mathrm{opt}}(\bm \gamma) \right) ^{2}}{2 \sigma_{q_{\mathrm{w}}}^{2}}-\dfrac{\left( T_{\mathrm{w}}^{i, \mathrm{meas}}-T_{\mathrm{w}}^{i, \mathrm{opt}}(\bm \gamma) \right) ^{2}}{2 \sigma_{T_{\mathrm{w}}}^{2}}\right],
\end{split}
\end{equation}
where the dependence of the optimal values on the two material properties has been made explicit for clarity. 

Given $\bm M$ and a value for the couple of catalytic coefficients, the optimal nuisance parameters and associated heat fluxes are determined using the BL code. The procedure for this optimization is the Nelder-Mead algorithm~\cite{Nelder}, which is a gradient-free method requiring only evaluations of the BL model solution. Typically, a few hundreds resolutions of the BL model are needed to converge to the optimum of~\eqref{eq:15}. The computational cost of the optimization prevents us from using directly this approach to draw samples of $\bm \gamma$ from their posterior distribution, and this fact motivates the approximation of the optimal (log) likelihood in~\eqref{eq:17} as discussed in Subsection~\ref{Surrogates}.

\subsubsection{Prior distributions}
To complete the Bayesian formulation, we now discuss the selection of the prior for the catalytic coefficients $\bm \gamma$.
We start by observing that, although it was assumed that the reference probe is well characterized when designing the two-probe experiment, the two coefficients $\gammar$ and $\gammap$ play a similar role in the expression of the likelihood in~\eqref{eq:17}. In fact, the observations should contribute to learn about both material properties. In other words, the differences in the knowledge of $\bm \gamma$ should be reflected by their distinct priors and not in the design of the likelihood. Therefore, it is important to select priors that fairly account for the initial beliefs in the values of the catalytic coefficients. 
In this case, we have to be cautious with our choice. Considering first the catalytic property of the reference calorimeter, previous works~\cite{Greaves, Greaves2, Young, Dickens, Hartunian, Myerson, Melin, Cauquot, Meyers, Driver} show that the a priori knowledge of $\gammar$ is actually quite poor: values proposed in literature vary significantly from one experiment to another. Furthermore, $\gammar$ has been reported for a limited number of conditions, leaving us with large prior uncertainties since in our experiment the boundary layer edge conditions are unknown too. Similarly, the initial knowledge of $\gammap$ is poor. For instance, previous works (\textit{e.g.}~\cite{Panerai}) show that the value of $\gammap$ can span two orders of magnitude depending on the testing conditions. To conclude, constructing a sharp prior distribution for $\gammap$ on the basis of previous works is difficult, while assuming a better knowledge of $\gammar$ is not realistic. For all these reasons, we decided in this work to consider independent priors with initial ranges spanning few orders of magnitude, stating bounds on plausible values:
$$
    10^{-4} \le \gammar , \gammap \le 1.
$$
The lower and upper bounds were set to encompass values proposed in the literature and to ensure that they contain the values to be inferred.
Based on the proposed bounds, the last step to derive the prior consists on specifying the distribution withing the range. Here, instead of using an non-informative prior where any value is as likely as any other (\textit{i.e.}, a uniform prior), we decided to go for log-uniform distributions, 
$$
    \log_{10}(\gammap), \log_{10}(\gammar) \sim \mathcal U(-4,1),
$$
which are better suited when the priors range over several orders of magnitude. 

The theoretical models describing the chemically reacting boundary layer, together with the experimental data available are integrated in the Bayesian framework for the inference of $\bm \gamma$. In the next subsection, we describe how we reduce the computational complexity inherent to the sampling of the posterior. Specifically, we rely on a surrogate model for the log-likelihood function to alleviate most of the computational burden.

\subsection{Surrogate model for the log-likelihood function}
\label{Surrogates}

The log-likelihood function must be evaluated multiple times when sampling the posterior distribution using MCMC methods. Since an evaluation of the log-likelihood requires many resolutions of the reacting boundary layer model to determine the optimum boundary layer edge conditions, direct sampling strategies based on the full model would be too costly. 
To overcome this issue, the logarithm of the likelihood function~\eqref{eq:17} is approximated by a surrogate model whose evaluations are computationally cheap.

\subsubsection{Parametrization}
To construct a surrogate model of the log-likelihood, we first introduce new canonical random variables, $\bm \xi = (\xi_1,\xi_2)$, for the parametrization of the catalytic coefficients. We set $\bm \xi$ to be uniformly distributed over the unit square: $\bm \xi \sim \mathcal U [0,1]^2$.
Then, we fix $\gammap(\xi_1) = 10^{-4\xi_1}$ and $\gammar(\xi_2) = 10^{-4\xi_2}$, such that $\gammap(\bm \xi)$ and $\gammar(\bm \xi)$ are independent, identically distributed, and follow log-uniform distributions with range $[10^{-4},1]$. 
The Bayesian inference problem can finally be recast in terms of the canonical random variables, leading to 
\begin{equation} \label{eq:18a}
P(\bm \xi \vert \bm M) \propto L^{\mathrm{opt}}(\bm M \vert \bm \gamma(\bm \xi)) P (\bm \xi), \quad
P (\bm \xi) = \begin{cases} 1, & \bm \xi \in [0,1]^2, \\
0, & \mbox{otherwise}. 
\end{cases}
\end{equation}
We seek to construct a surrogate of the optimal likelihood $L^{\mathrm{opt}}(\bm M \vert \bm \gamma(\bm \xi))$ with this parametrization. In particular, we decided to proceed with the log-likelihood instead of the likelihood as it ensures the positivity of the approximation. More precisely, we aim for a surrogate model of $Y(\bm \xi)$ defined by
$$
Y(\bm \xi) \doteq \log \left(L^{\mathrm{opt}}(\bm M \vert \bm \gamma(\bm \xi)) \right).
$$
Below, we propose to use a Gaussian Process (GP) model to approximate $Y(\bm \xi)$.

\subsubsection{Gaussian process model}\label{sec:gp}
GP models ~\cite{GP} have been widely used in uncertainty propagation, sensitivity analysis, optimization and inverse problems~\cite{Chen}. Due to their statistical nature, a GP provides a measure of the uncertainty (variance) in the prediction. The main premises of the GP lay in the assumption that the function to be approximated is the realization of a Gaussian process characterized by its mean $\mu(\bm \xi)$ and two-point covariance $C_{\mathrm{GP}} (\bm \xi,\bm \xi')$ function. Then, from the observation of the function values $Y^{(i)}$ at the sample points $\bm\xi^{(i)}$, one can derive the posterior distribution of the GP model~\cite{GP}, and evaluate the GP mean and variance at any new point $\bm \xi$. The selection of the prior of the GP model is a crucial step. In this work we tested several zero-mean, stationnary processes with covariance functions from the Matern's class~\cite{Matern}; we found that the log-likelihood function is well approximated using the standard isotropic squared exponential kernel, given that both catalytic parameters play the same role in the likelihood. The covariance function then reads
\begin{equation} \label{eq:19}
C_{\mathrm{GP}}({\bm\xi}, \bm{\xi'}) = \sigma^{2}_{\mathrm{GP}} \exp \left(-\dfrac{1}{2L^{2}_\mathrm{GP}} (\bm\xi - \bm\xi')^{\mathrm{T}}(\bm\xi - \bm\xi')\right),
\end{equation}
where $L_{\mathrm{GP}}$ and $\sigma_{\mathrm{GP}}^2$ are the a priori correlation length and variance of the GP. All results presented hereafter use the covariance function in~\eqref{eq:19}. Denoting $\bm Y = (Y^{(1)} \cdots Y^{(p)})^{\mathrm{T}}$ the vector of observations, the posterior mean of the GP model, or the best prediction of $Y(\bm \xi)$ is
\begin{equation} \label{eq:20}
\Esp{Y_{\mathrm{GP}} (\bm\xi)} = \bm k^{\mathrm{T}} (\bm\xi) {\bm K}^{-1} \bm Y,
\end{equation}
where the vector $\bm k(\bm \xi)$ and matrix $\bm K$ are given by
$$
    \bm k_i(\bm \xi) = C_{\mathrm{GP}}(\bm \xi,\bm \xi^{(i)}), \quad
    \bm K_{i,j} = C_{\mathrm{GP}}(\bm \xi^{(i)},\bm \xi^{(j)}) + \sigma_\epsilon^2 \delta_{i,j},
$$
where $\delta_{i,j}$ is the Kronecker symbol. The variance of the prediction is
$$
    \Var{ Y_{\mathrm{GP}} (\bm\xi)} = 
    C_{\mathrm{GP}} (\bm \xi,\bm \xi) - \bm k^{\mathrm{T}} (\bm\xi) {\bm K}^{-1} \bm k (\bm\xi).
$$
In the expression of the matrix $\bm K$ above, $\sigma_\epsilon$ corresponds to the error on the estimation of $Y^{(i)}$. As $\sigma_\epsilon \rightarrow 0$ we have $\Esp{Y_\mathrm{GP} (\bm \xi^{(i)})} \rightarrow Y^{(i)}$, while $\Var{Y_{\mathrm{GP}} (\bm\xi^{(i)})}$ vanishes at all observation points.
In practice $\sigma_\epsilon$ is set to a small but non-zero value to ensure that the matrix $\bm K$ is invertible. Concerning the parameters of the covariance function, $\sigma_{\mathrm{GP}}^2$ and $L_{\mathrm{GP}}$, they are inferred from the observations through the maximization of their resulting joint log marginal likelihood~\cite{GP}.


\subsection{Posterior sampling}\label{mcmc}
\begin{figure}[hbt!]
\centering
\includegraphics[width=0.8\textwidth]{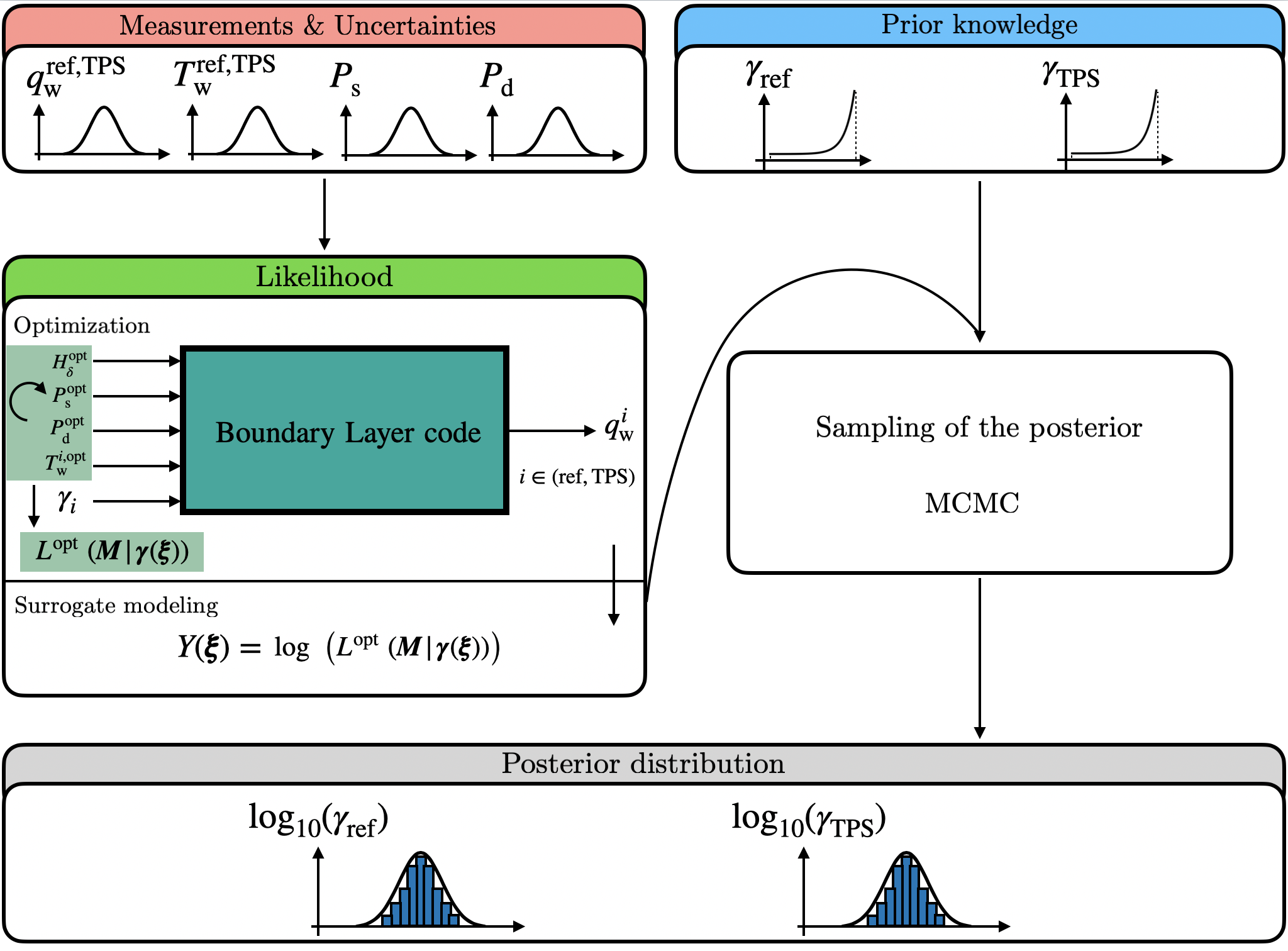}
\caption{Bayesian inference framework in a nutshell}
\label{fig:5}
\end{figure}

Sampling the posterior distribution of $\bm \xi$, and so generate samples of the catalytic coefficients, is made possible thanks to the construction of a surrogate model which is very cheap to evaluate. To sample from the posterior distribution we use the Metropolis-Hastings MCMC algorithm. The algorithm consists in a sequence of steps, where a random move from the current step $\bm \xi_{i}$ to $\bm \xi^*$ is proposed. Denoting $r \doteq P(\bm \gamma(\bm \xi^*)\vert M)/P(\bm \gamma(\bm \xi)\vert M)$ the ratio of the posteriors, the move is accepted ($\bm \xi_{i+1} = \bm \xi^*$) with a probability $\min(1,r)$, otherwise $\bm \xi_{i+1} = \bm \xi_{i}$. The algorithm used in the present work relies on Gaussian increments $\bm \xi^* - \bm \xi_{i}$. The covariance matrix of the proposal distribution is defined through
\begin{equation} \label{eq:Ct}
\bm C_{\mathrm{Prop}} = s \ \bm{C}_\mathrm{Post},
\end{equation}
where $\bm{C}_\mathrm{Post}$ is the posterior covariance of $\bm \xi$ and $s = 2.38^2/d$ is a scaling factor dependent on the dimension of the sampling space (here $d=2$). Because the posterior covariance of $\bm \xi$ is not known, it is progressively estimated from the samples drawn during  an initial ``burn-in'' stage of chain~\cite{Kaipio}. 
The scaling factor is selected to ensure an acceptance rate varying between 20 and 50\% by following Roberts et al~\cite{Roberts}, and ensure a sufficiently fast decorrelation of the chain. Figure~\ref{fig:5} illustrates the different elements constituting our Bayesian framework.

\section{Case study}\label{Results}

The methodology presented in the previous sections is used for a real case of plasma wind tunnel testing. This case is used to assess the validity and possible shortcomings of the approach.

\subsection{Experimental data and associated uncertainties}
\label{Results}

The experimental run used for this case study is depicted in Table \ref{tab:data_S1}. Uncertainties are involved in two different processes. The most natural kind of uncertainty that we can characterize is intrinsic to the measurement device and its accuracy at measuring. On top of this, there is the measuring uncertainty, where, ideally, one could perform a measurement infinite times and perform statistics on that, extracting the relevant parameters of the resulting distribution. As we do not have infinite time or resources, a t-student distribution is assumed and the sample Gaussian is corrected by the t-factor. Overall, we account for both sources of uncertainties as the squared sum. It is important to remark that the quantities considered for both $q_{\mathrm{w}}^{\mathrm{ref}}$ and $q_{\mathrm{w}}^{\mathrm{TPS}}$ are derived from the measurement of the material emissivities as already explained in Sec.~\ref{experiments}.

\begin{table}[hbt!]
\centering
\caption{Experimental data and uncertainties considered in our case study. Data taken from \cite{Panerai}} 
\label{tab:data_S1}
  \begin{tabular}{ c  c  c  c  c  c  c }
    \toprule
    \textbf{Experiment $\mathrm{S_{1}}$} & $q_{\mathrm{w}}^{\mathrm{ref}} \ \mathrm{[kW/m^{2}]}$ & $T_{\mathrm{w}}^{\mathrm{ref}} \ \mathrm{[K]}$ & $\Ps \ \mathrm{[Pa]}$ & $\Pst \ \mathrm{[Pa]}$ &  $q_{\mathrm{w}}^{\mathrm{TPS}} \ \mathrm{[kW/m^{2}]}$ & $T_{\mathrm{w}}^{\mathrm{TPS}} \ \mathrm{[K]}$\\ \midrule
    Mean ($\mu$)& 195 & 350 & 1300 & 75 & 91.7 & 1200\\
    Std deviation ($\sigma$)& 6.5& 11.7 & 1.3 & 1.5 & 3.05 & 40\\ \bottomrule

  \end{tabular}
\end{table}

\subsection{Log-likelihood approximation}
\label{loglikelihood}

\begin{figure}[!htp]
\centering
\includegraphics[width=0.6\textwidth]{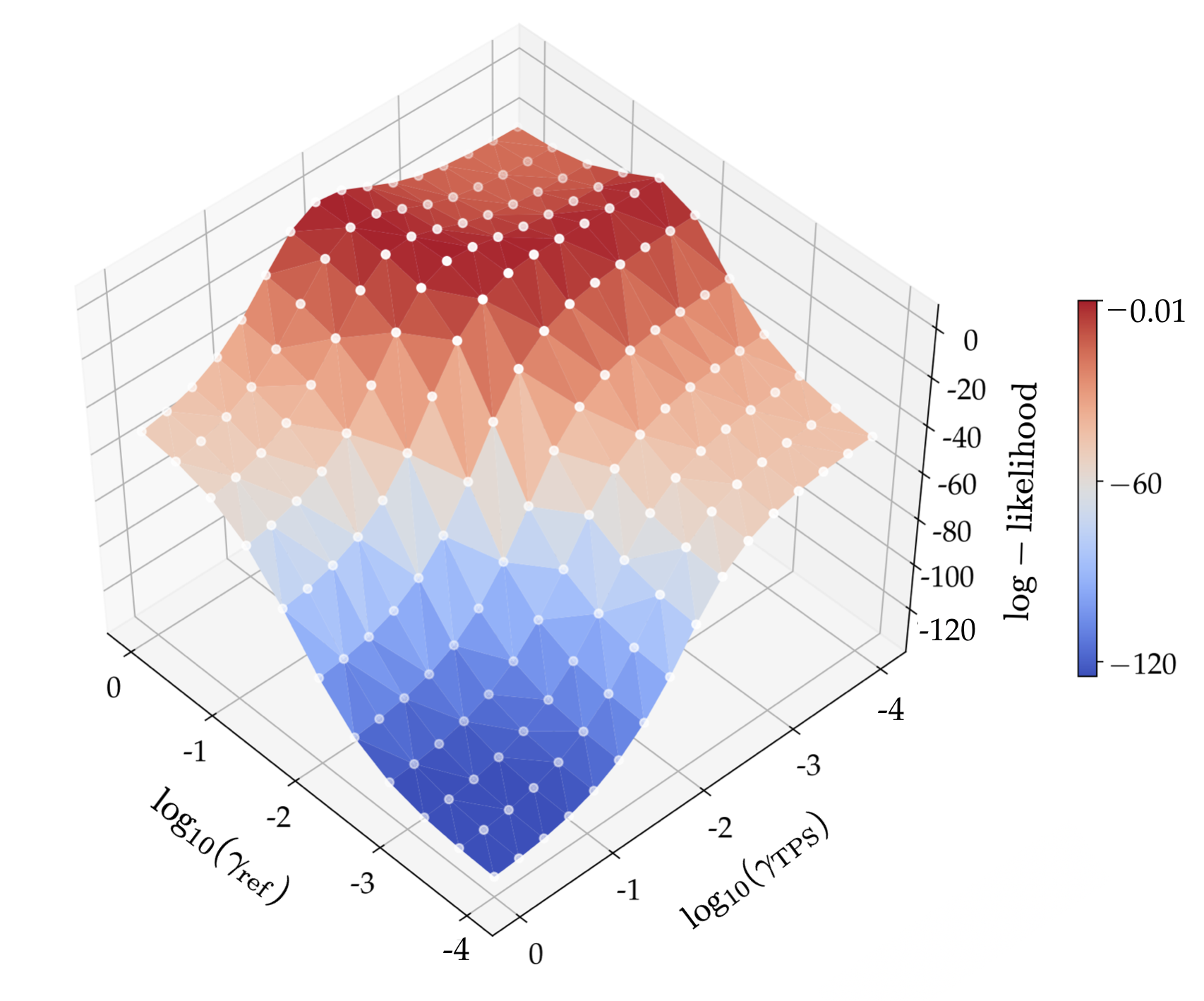}
\caption{Log-likelihood function evaluated on the chosen $\mathrm{\gamma_{ref}, \gamma_{TPS}}$ grid}
\label{fig:7}
\end{figure}

We run the optimization algorithm in a uniform grid of 176 points on the space of the $\mathrm{log_{10} (\gamma_{ref})}$ and $\mathrm{log_{10} (\gamma_{TPS})}$ variables. This uniform grid is chosen slightly asymmetric, 11x16 points, repectively. This choice for an initial grid gives us better refinement on the $\mathrm{log_{10} (\gamma_{TPS})}$ direction. From physical considerations, we expect the variability of this parameter to be greater on the posterior than $\mathrm{log_{10} (\gamma_{ref})}$. Therefore, providing a more refined grid on $\mathrm{log_{10} (\gamma_{TPS})}$ could produce better approximations for less cost than a squared grid of that size. Fig. \ref{fig:7} shows the log-likelihood function evaluated at these grid points. These evaluations are then used to construct the surrogate approximation by transforming the physical variables $\mathrm{log_{10} (\gamma_{ref})}$ and $\mathrm{log_{10} (\gamma_{TPS})}$ into their respective canonical counterparts $\bm \xi$ as explained in Sec.~\ref{Surrogates}.

Overall, the shape of the log-likelihood function falls from the compatibility of the given pair of $\gammar$ and $\gammap$ with the observed quantities measured in the plasma wind tunnel. In general, for large values of $\gammar$, log-likelihood values tend to be larger. The same happens with $\gammap$ for low values. This is already hinting at the fact that higher catalytic activity is expected for the reference material than for the protection material in question, for the given boundary layer edge conditions. On top of that, there is a range of values for $\gammar$ and $\gammap$ that represent the best agreement with the experimental data. This set of values have to be interpreted jointly: for high $\gammar$ values, $\gammap$ can only take values in a narrow range placed at the middle of its spectrum. For low values of $\gammar$ (mid-spectrum), $\gammap$ can take down to the minimum value of $\mathrm{10^{-4}}$. For large values of $\gammar$ and low values for $\gammap$, the heat flux, which is the quantity in the likelihood sensitive to our choice of catalytic parameters, is not sensitive enough to changes in those specific ranges. We explore later in Sec.~\ref{Preliminary} how this limitation set by the physical model can be overcome by modifying the experimental methodology. 

\begin{figure}[!htb]
\centering
\includegraphics[width=0.6\textwidth]{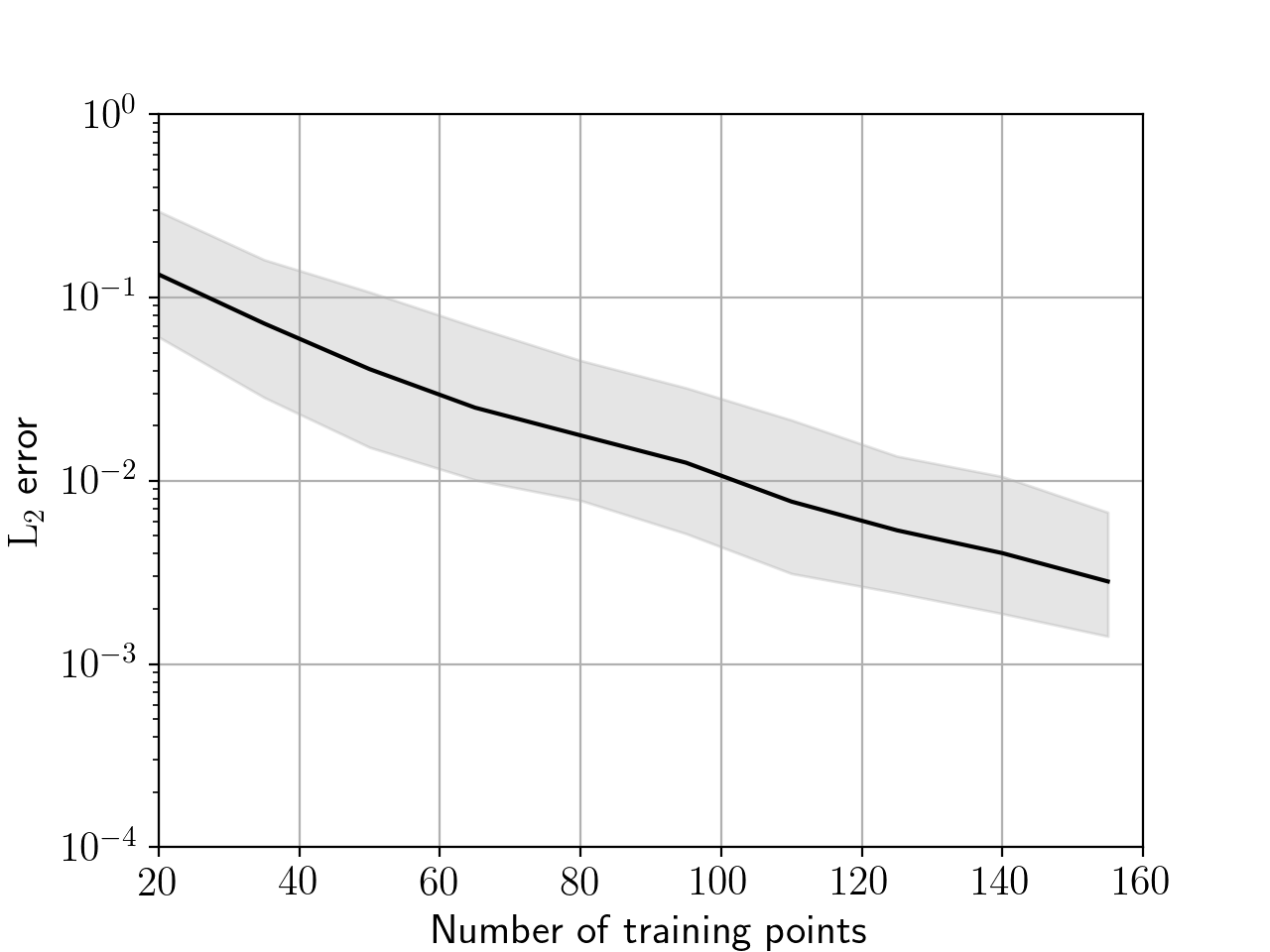}
\caption{Normalized $\mathrm{L_{2}}$ error norm of the GP approximation with varying number of training points}\label{fig:GP_L2}
\end{figure}

\begin{figure}[!hbt]
\centerline{
\begin{tabular}{|c|c|}\hline
GP mean 	& GP mean profile \cr \hline
\includegraphics[width=0.45\textwidth]{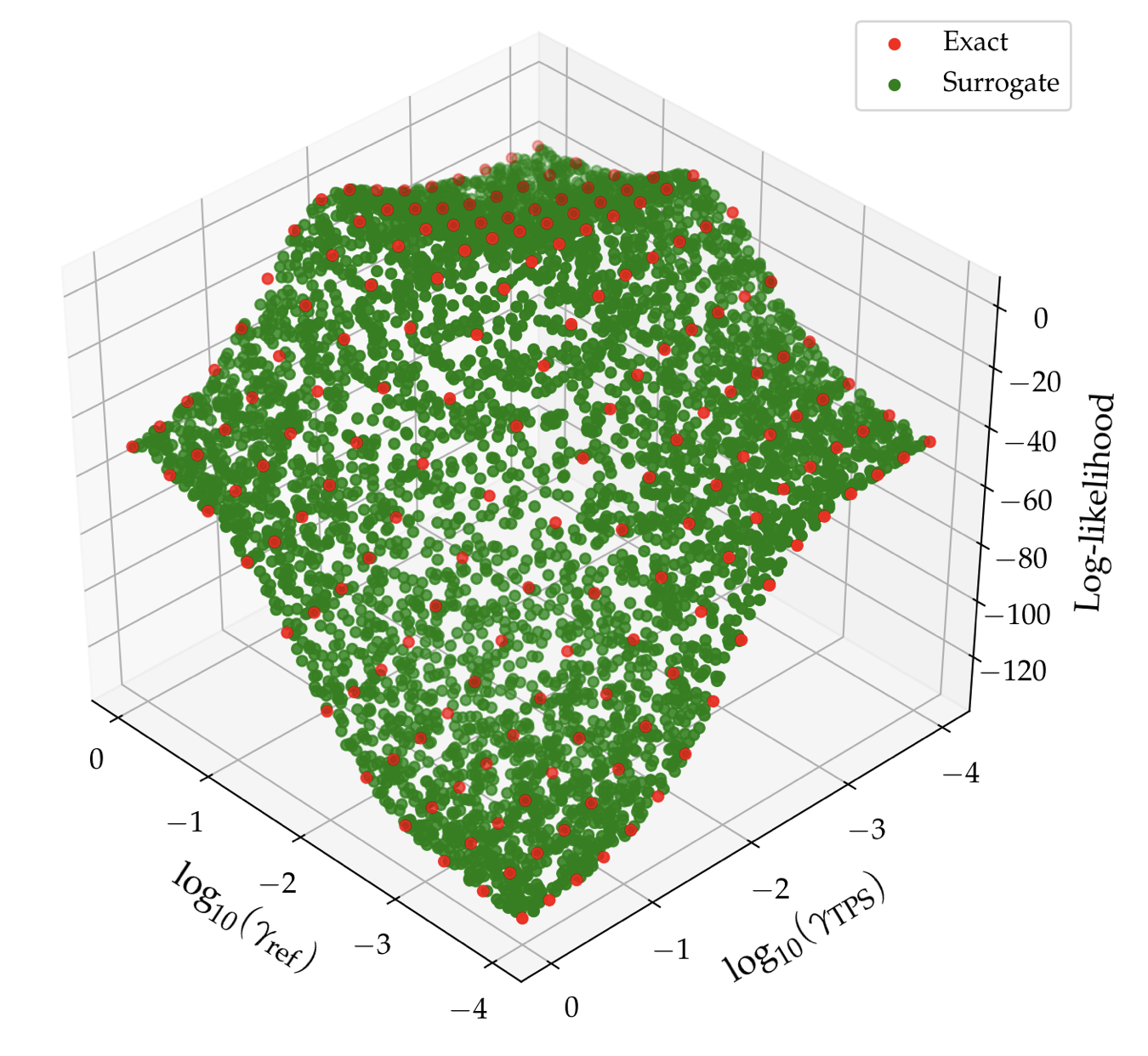} &
\includegraphics[angle=-0.5,width=0.45\textwidth]{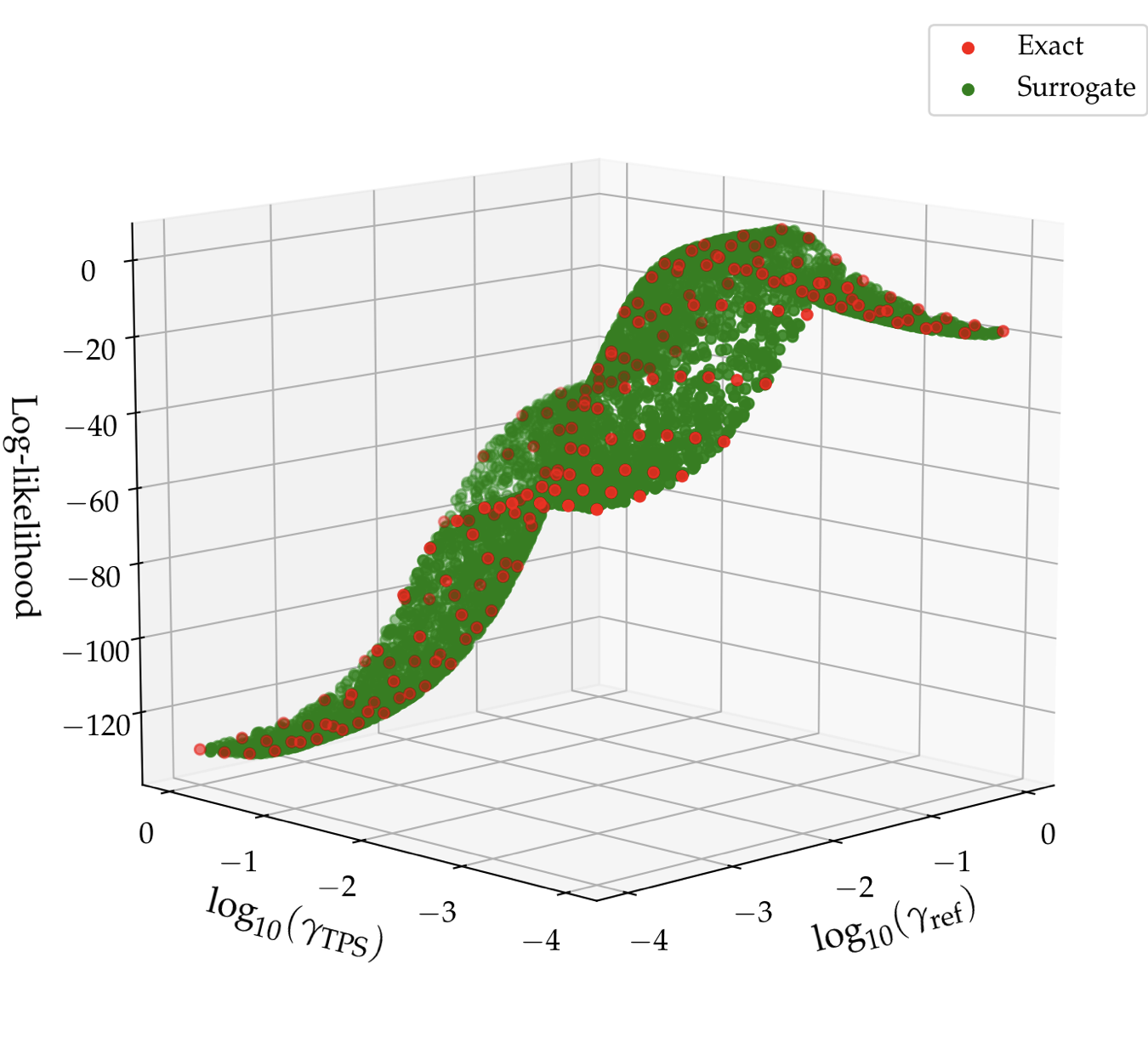} \cr \hline
\end{tabular}}
\caption{GP surrogate comparison with the exact log-likelihood values in logarithmic variables}\label{fig:9}
\end{figure}

As already mentioned previously in Sec. \ref{Surrogates}, we need to properly capture all these features of the log-likelihood with a surrogate model. We propose to use a GP surrogate. One of the advantages of using GP is that it gives us an estimation of the predictive variance. Fig.~\ref{fig:GP_L2} shows the normalized $\mathrm{L_{2}}$ error norm for the GP surrogate on a validation set with 10\% of the available points plotted against the number of training points. We carry out this procedure 1,000 times with different validation sets each time. The results show the mean and the 95\% confidence interval of the computed error. The approximation falls below a 1\% error on the validation set as the number of training points gets closer to 160. In practice, we use all model evaluations (176) to construct our GP, knowing that the approximation is already good enough. Furthermore, this approximation obtained has a maximum predictive standard deviation of $\mathrm{1\%}$. Fig. \ref{fig:9} shows the apparent good agreement between the mean value predicted by the GP and the data points computed for the log-likelihood in Fig. \ref{fig:7}.

\subsection{Sampling of the posterior distribution}
\label{Results:MCMC}

\begin{figure}[!htb]
\centerline{
\begin{tabular}{|c|c|}\hline
1,000,000 steps 	& 15,000 steps close-up \cr \hline
\includegraphics[width=0.45\textwidth]{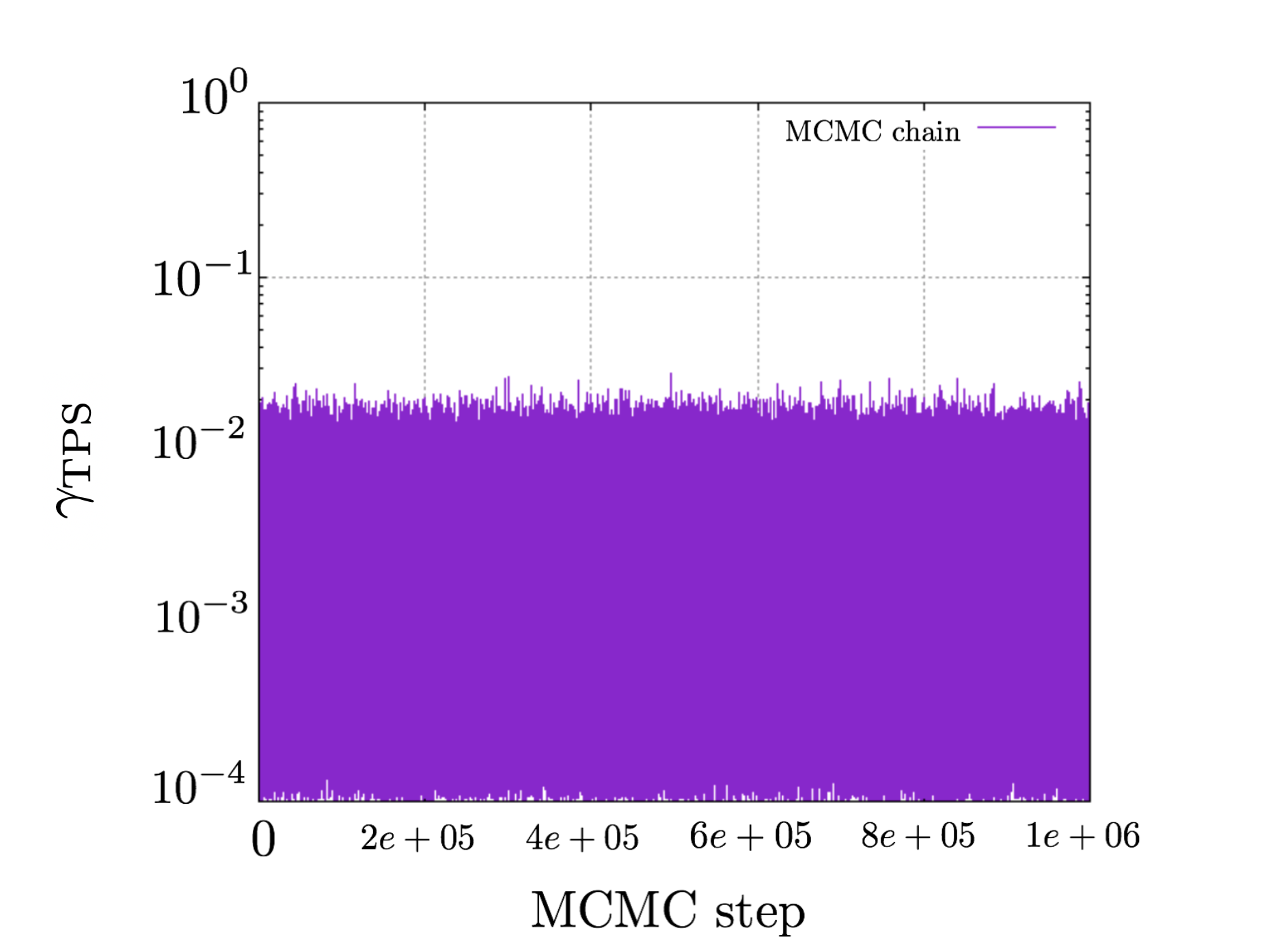} &
\includegraphics[width=0.45\textwidth]{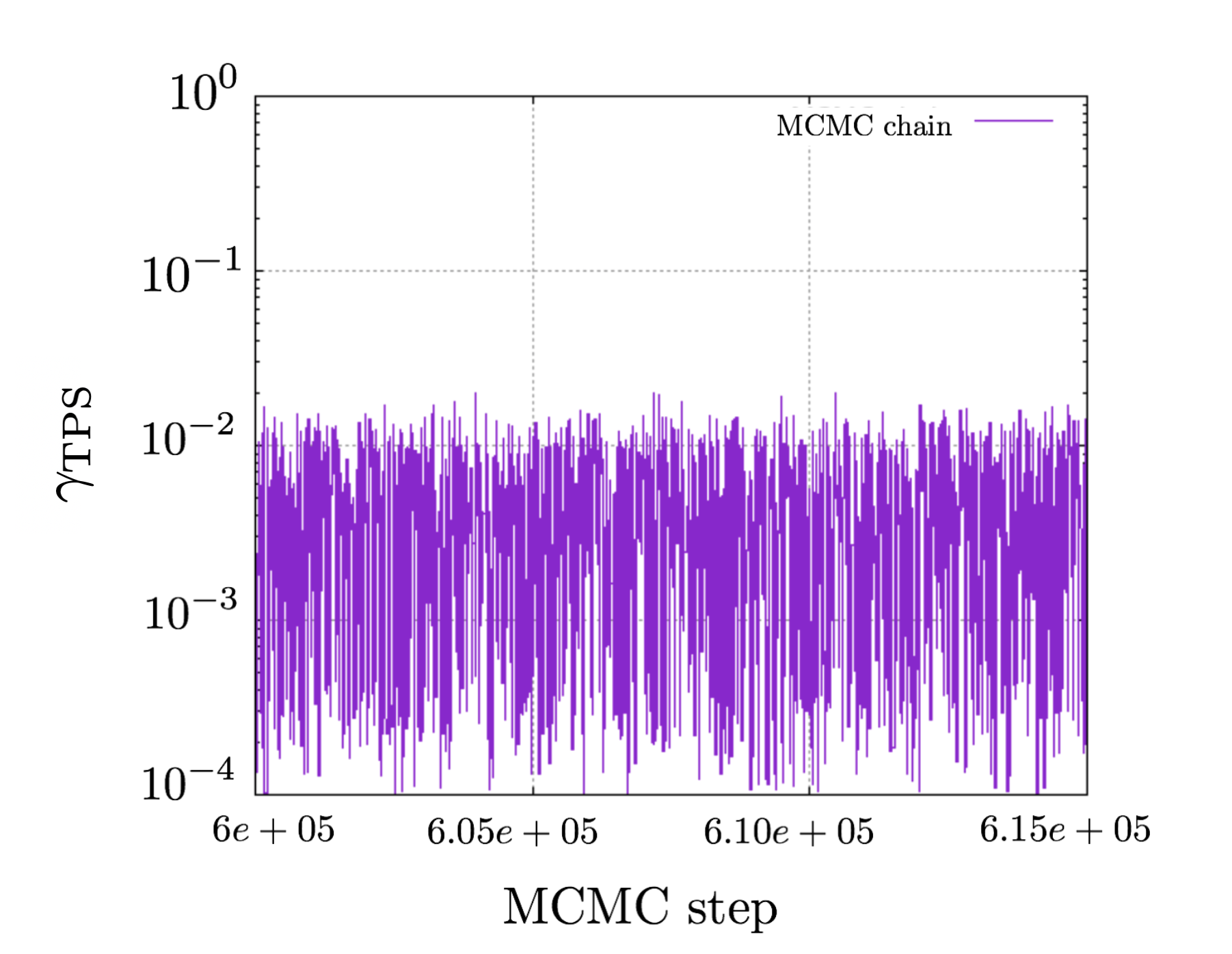} \cr \hline
\end{tabular}}
\caption{Chain obtained with 1,000,000 steps and 15,000 steps (right)}\label{fig:11}
\end{figure}

We perform a MCMC sampling for the choice of GP surrogate. The chains obtained are depicted in Fig. \ref{fig:11}. We can see that the chains present no long-term correlation and mix well.

\begin{figure}[!htb]
\centering
\includegraphics[width=0.57\textwidth]{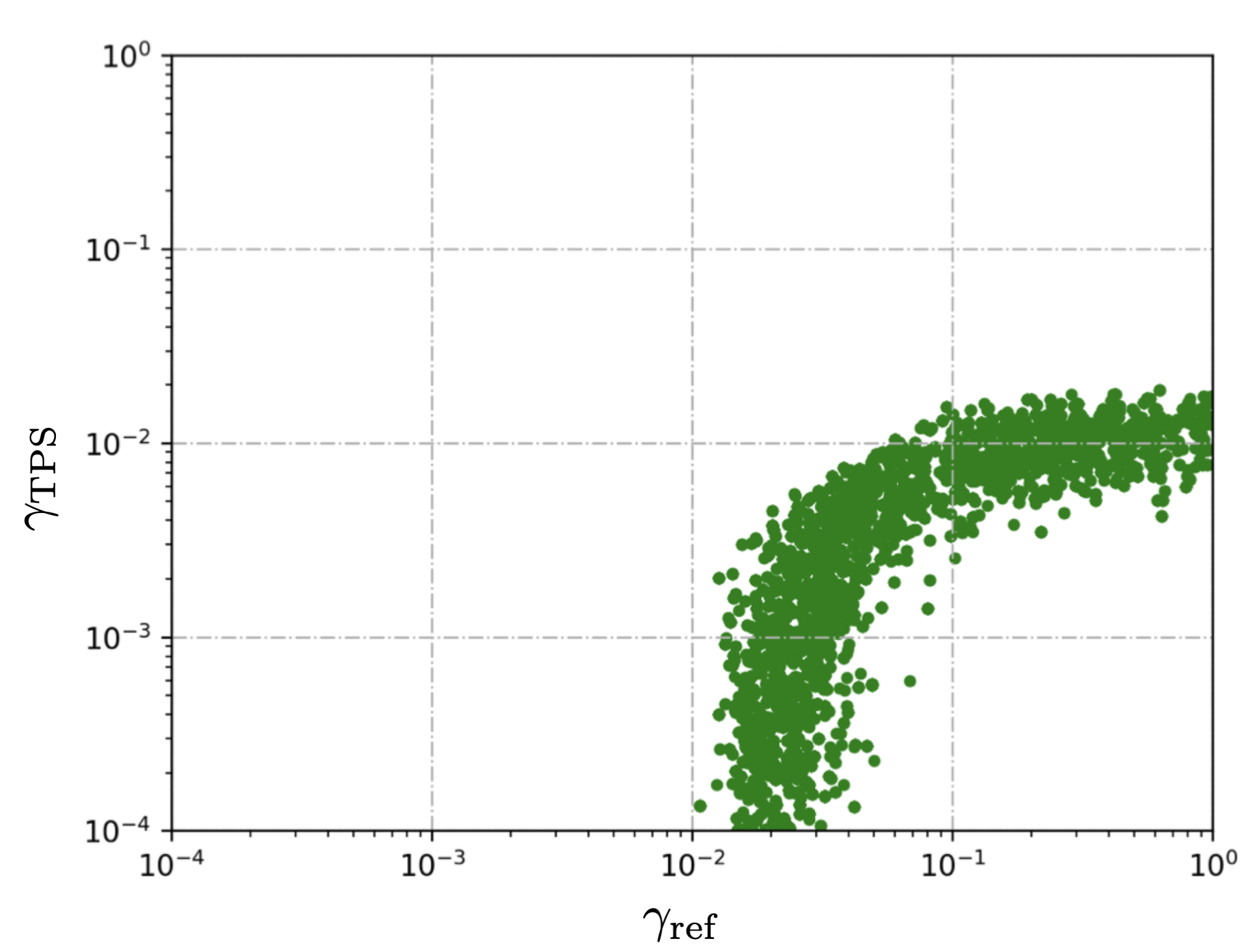}
\caption{Joint posterior samples of the MCMC algorithm}
\label{fig:GPChain}
\end{figure}

The posterior samples obtained are shown in Fig. \ref{fig:GPChain}. In general, the tendency of the samples is to remain in a narrow area of the $\gammap$ space when $\gammar$ takes large values. Once $\gammar$ starts moving towards lower values this tendency is reversed and $\gammap$ can take values in a wider range while $\gammar$ is confined in a narrow region. This joint behavior falls from the inference framework. The key variable here is the boundary layer edge enthalpy $\Hd$ which is shared between both materials tested (reference and TPS). When the model takes up large values for $\gammar$, a large amount of the observed heat flux for the reference material is explained in the model through the magnitude of this parameter, setting low the influence of the enthalpy $\Hd$. Low enthalpy needs larger $\gammap$ values to account for the observed heat flux on the protection material surface. The same happens for low values of $\gammar$ and $\gammap$. In this case, the values that lay interior to the shape defined by the posterior samples are not in agreement with observations for the reason just explained: large $\gammar$ needs large $\gammap$. The fact that ``large" and ``low" are also defined within a range (e.g not more than $\sim \mathrm{10^{-1.8}}$ for $\gammap$ and not less than $\mathrm{\sim 10^{-2}}$ for $\gammar$) is not imposed by the inference problem setting but by the physics-based model which makes some assumptions regarding the chemical nature of the flow. Some values of $\gammar$ and $\gammap$ could not explain, under the same $\Hd$, the observations. Overall, this behavior will naturally reflect on the marginal posterior distributions depicted in the next subsection.

\subsection{Discussion on the posterior distribution}
\label{Results}

The posterior marginals are reported below in Fig. \ref{fig:15}. We can observe that the distributions of both $\gammar$ and $\gammap$ drop to small values at both ends of the spectrum, reducing the support from the prior distributions proposed. This satisfying behavior can be explained by the proposed likelihood form, which uses all the available measurements to access the fitness of the model predictions. As a result, the formulation predicts that the values of $H_{\delta}^{\mathrm{opt}}$ that could explain the whole set of measured fluxes, temperatures and pressures, are actually far away from the maximum likelihood points when $ \gammar \ll 1$ and $ \gammap$ reaches large values. It is also important to notice that both distributions have well-defined peaks for $\mathrm{\gamma_{ref}\simeq 0.016}$ and $\mathrm{\gamma_{TPS}\simeq 0.01}$. The ranges of values observed in our calibration for both gammas fit perfectly with the model previously assumed by the experimentalists where the reference parameter takes higher values than the catalytic parameter of the protection material. It is also important to emphasize the fact that in this framework no assumptions are made concerning $\gamma_{\mathrm{ref}}$, which is estimated along with the protection material parameter with no differences in their prior knowledge. It can be suggested that a deeper experimental study can provide more insights to the behavior of the reference material and a different prior can be defined for the same analysis where differences in knowledge between the two probes can be then accounted for. 

\begin{figure}[!hbt]
\centering
\includegraphics[width=0.6\textwidth]{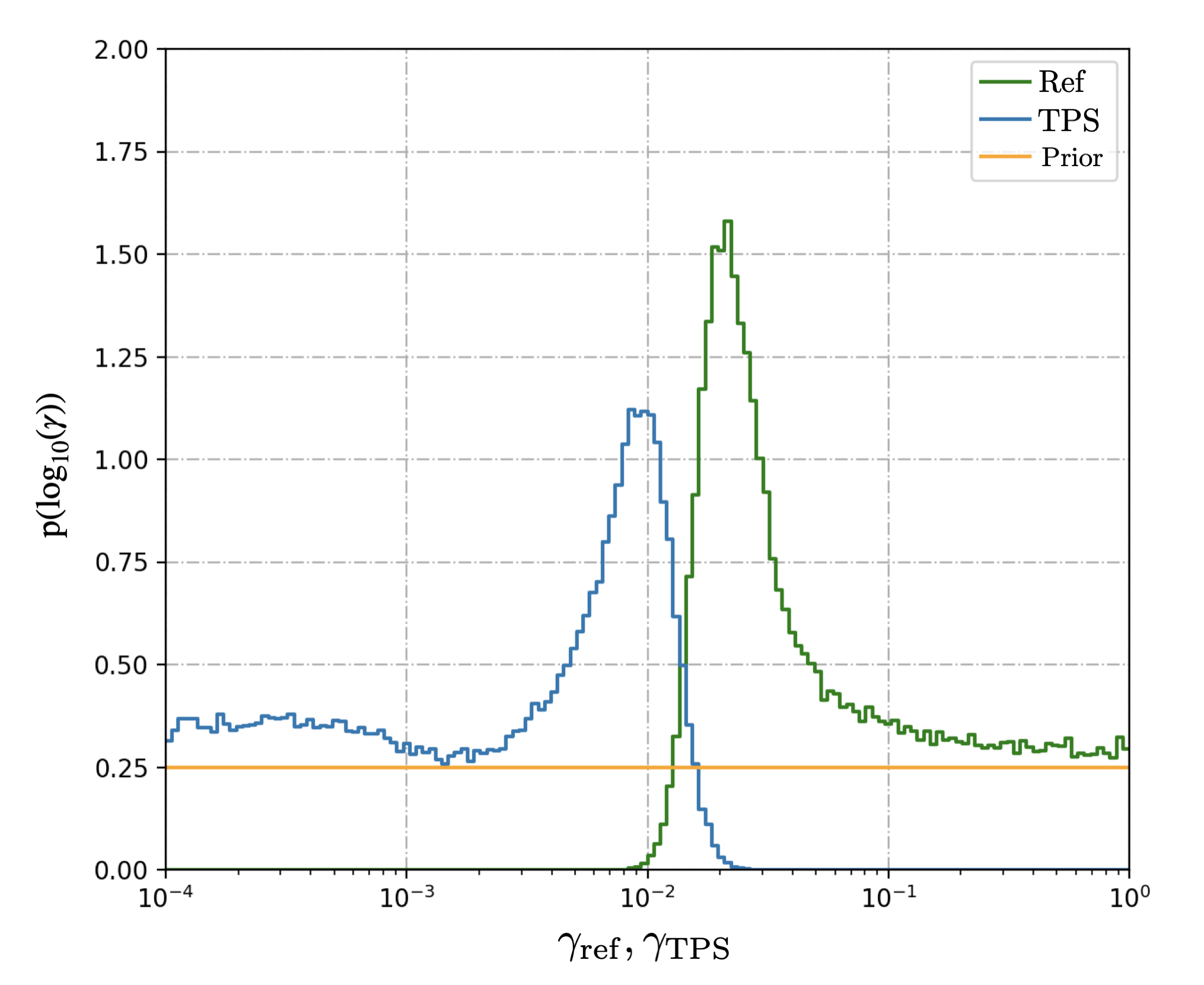}
\caption{Posterior marginals obtained for $\gammar$ and $\gammap$}
\label{fig:15}
\end{figure}

The statistics associated with these distributions are gathered in Table \ref{tab:table_stats}. The differences between these results and the outcomes of \cite{Sanson} are clear when looking at the values in the table and the shapes of the distribution functions obtained. A reduction of almost 20\% of the standard deviation and 40\% of the Coefficient of Variation (CV) for the catalytic parameter of the reference material is observed. There is no reporting of the posterior statistics for $\gammap$ in \cite{Sanson}. The capability of learning $\gammap$ from experimental data is lost without any particular treatment of the nuisance parameters in the formulation of the inference problem. 



\begin{figure}[hbt!]
\centerline{
\begin{tabular}{c c}
\includegraphics[width=0.4\textwidth]{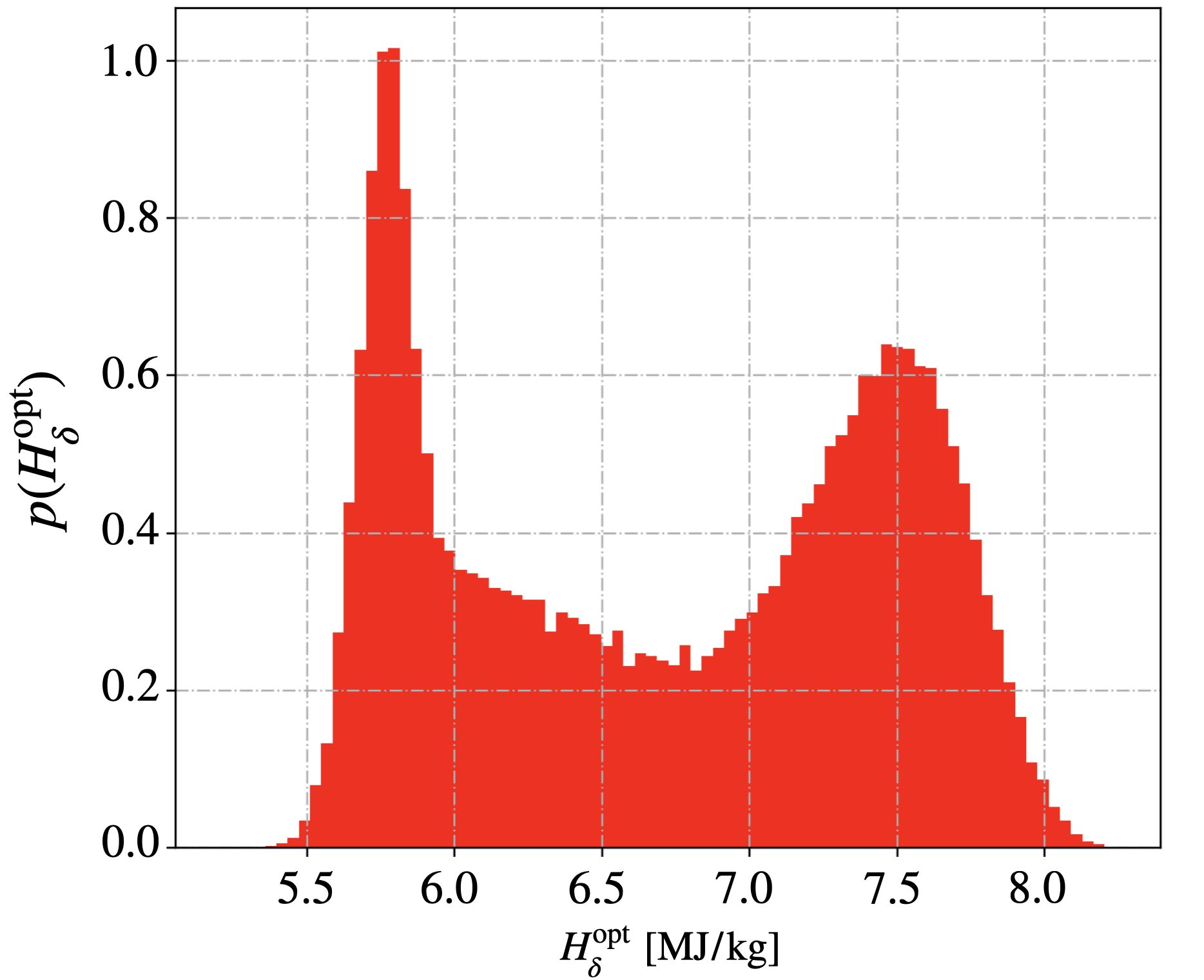} &
\includegraphics[width=0.41\textwidth]{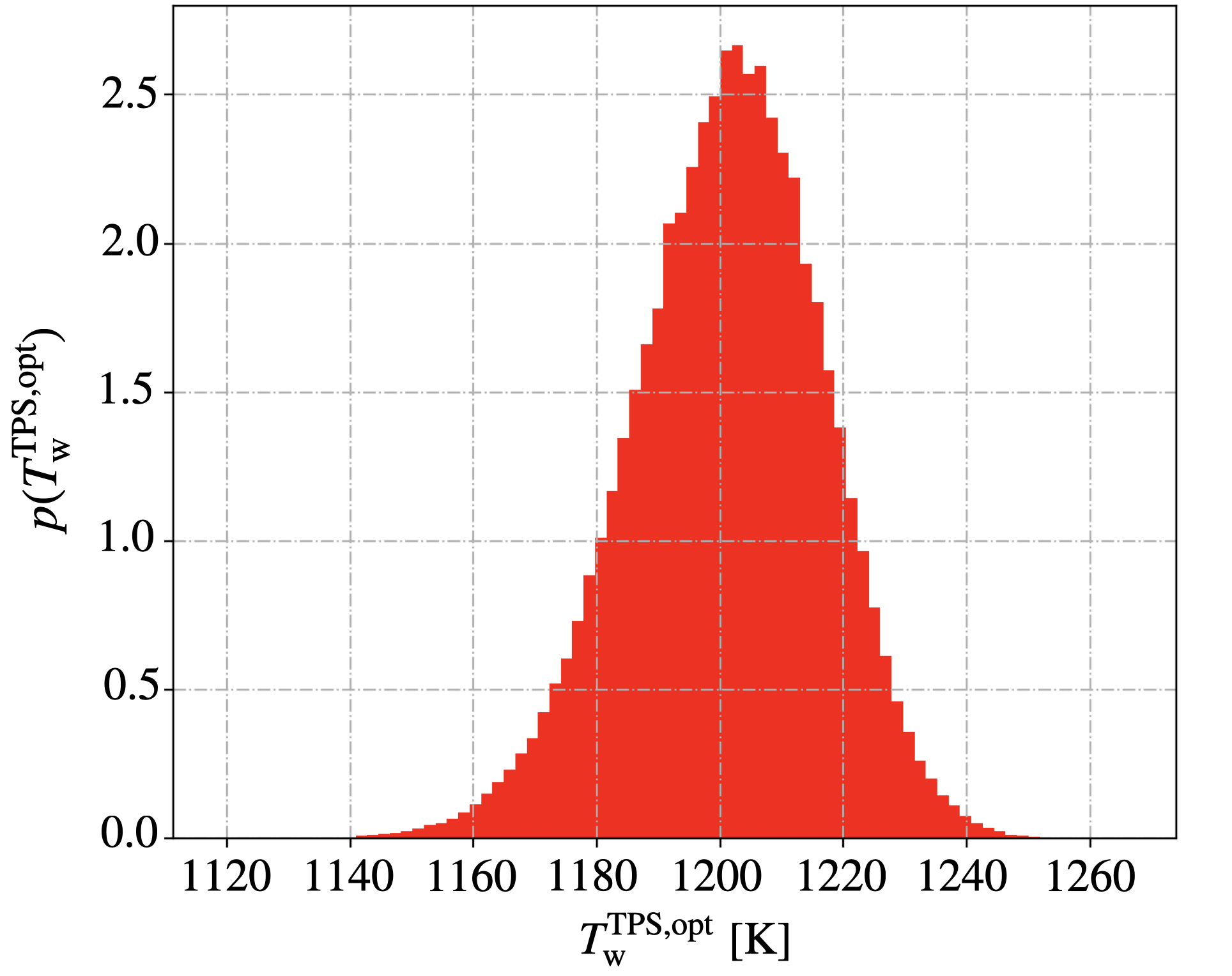} \cr
\includegraphics[width=0.4\textwidth]{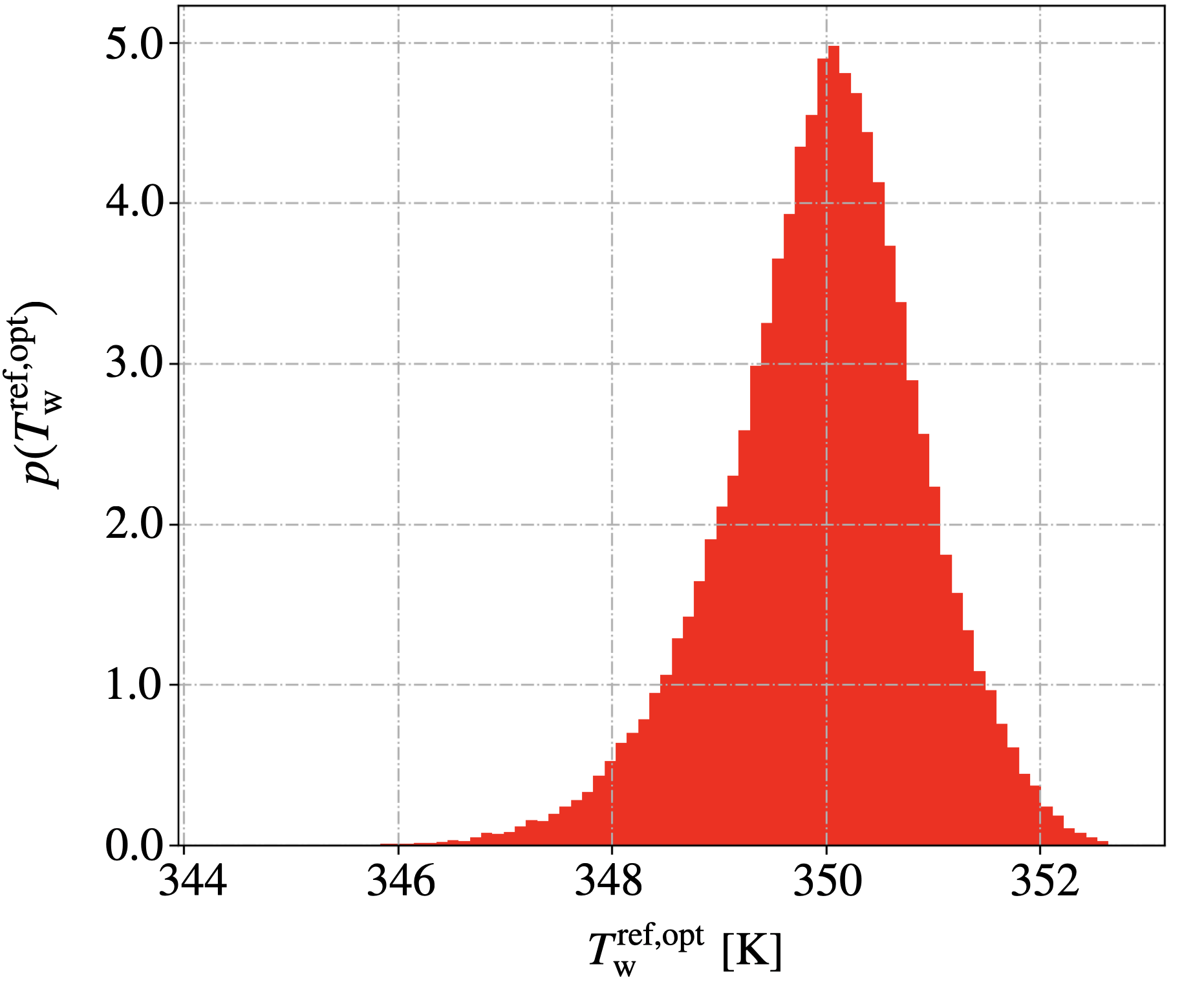} &
\includegraphics[width=0.41\textwidth]{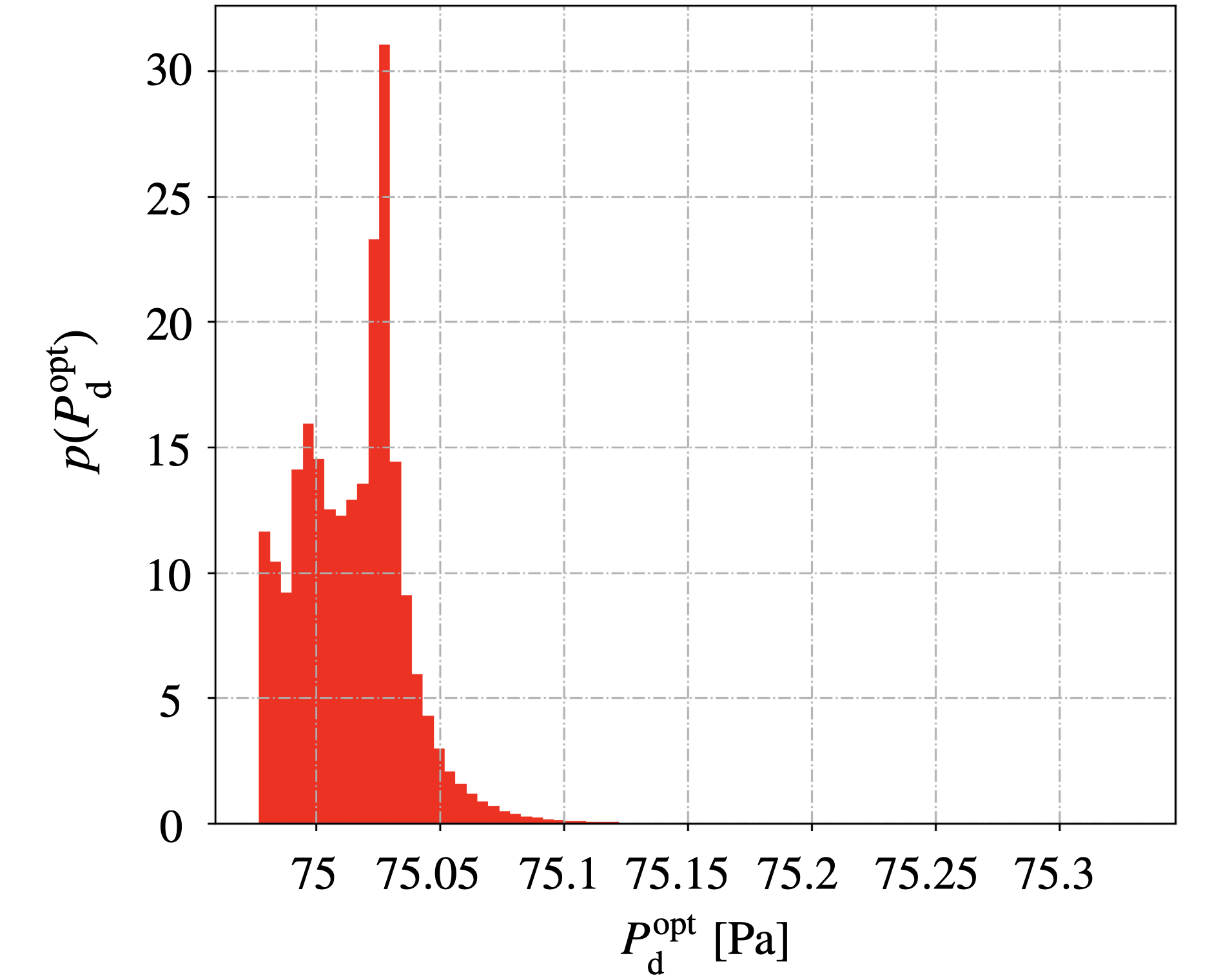}
\end{tabular}}
\caption{Distributions of the optimal nuisance parameters after propagating the posterior of $\gammar$ and $\gammap$}\label{fig:16}
\end{figure}

\begin{table}[!hbt]
\centering
\caption{Comparison of the posterior statistics for experiment $\mathrm{S_{1}}$ with the work of \cite{Sanson}} 
\label{tab:data}
  \begin{tabular}{ c  c  c  c  c }
    \toprule
     \textbf{Experiment} $\mathrm{\bm S_{1}}$ & Mean ($\mu$) & Std dev. ($\sigma$) & Max. A Posteriori (MAP) & CV [$\sigma/\mu$] \\ \midrule
    $\gammar$ & 0.060 & 0.078 & 0.022 & 1.3 \\
    $\gammap$ & 0.0034 & 0.0047 & 0.008 & 1.4 \\
    \textbf{Experiment} $\mathrm{\bm S_{1}}$ from \cite{Sanson} \\
    $\gammar$ & 0.042 & 0.095 & 0.018 & 2.3 \\
    $\gammap$ & - & - & - & - \\ \bottomrule

  \end{tabular}
  \label{tab:table_stats}
\end{table}
The distributions of the optimal parameters are also computed. For each of these quantities, a GP surrogate is computed on the same $\gamma_{\mathrm{ref}}, \gamma_{\mathrm{TPS}}$ grid than the log-likelihood. The resulting posterior samples of $\gammar$ and $\gammap$ are used as input for these surrogates, obtaining the distributions of the optimal parameters shown in Fig. \ref{fig:16}. A bimodal distribution is obtained for the enthalpy $H_{\delta}^{\mathrm{opt}}$. The shape of this distribution is a direct result from the optimization algorithm that computes the $H_{\delta}^{\mathrm{opt}}$, where many of its best points (``best" meaning the ones which maximize the likelihood) fall into two different group of values, decreasing the probability density among them. 

\begin{figure}[hbt!]
\centerline{
\begin{tabular}{c c}
\includegraphics[width=0.48\textwidth]{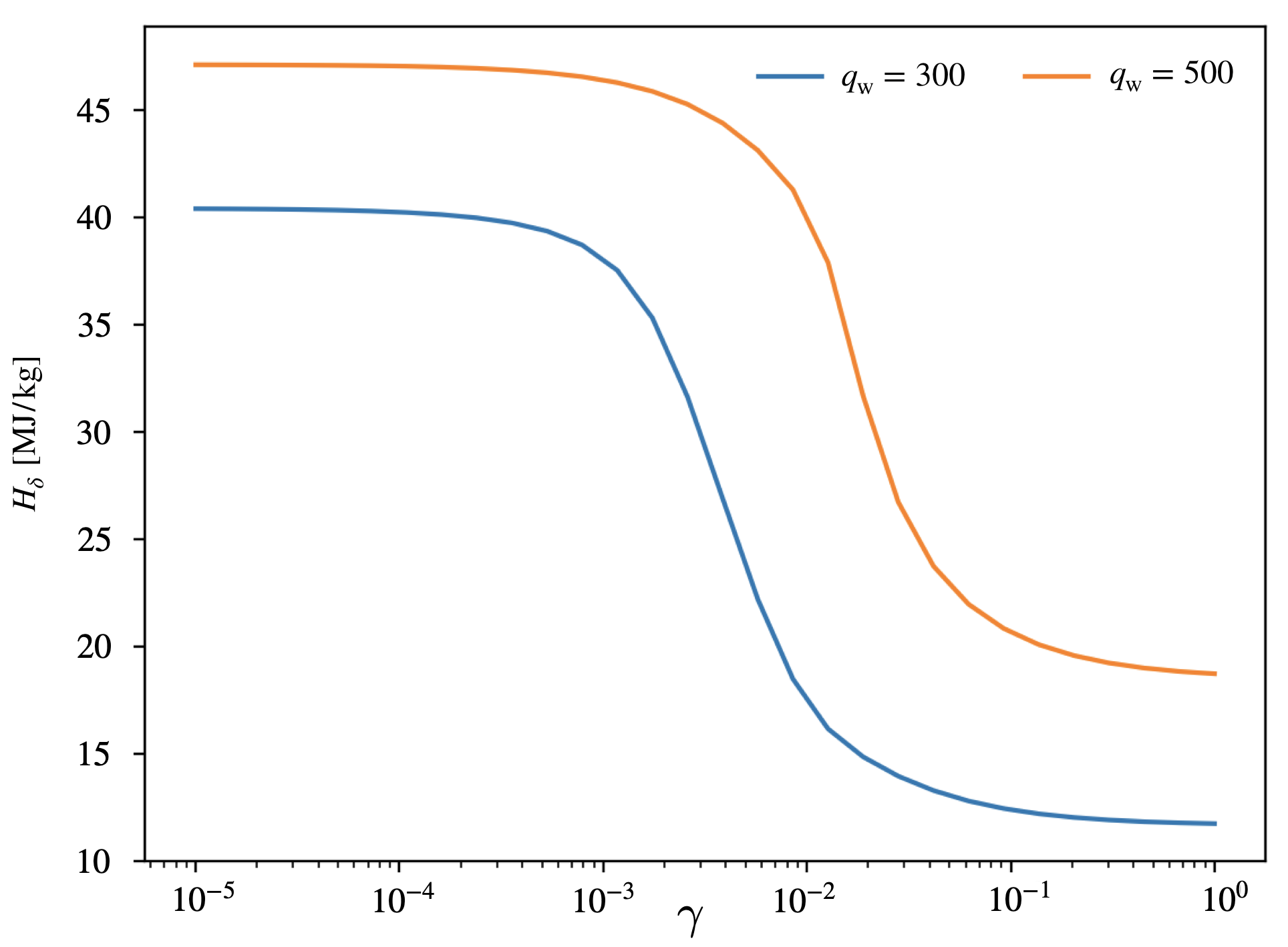} &
\includegraphics[width=0.48\textwidth]{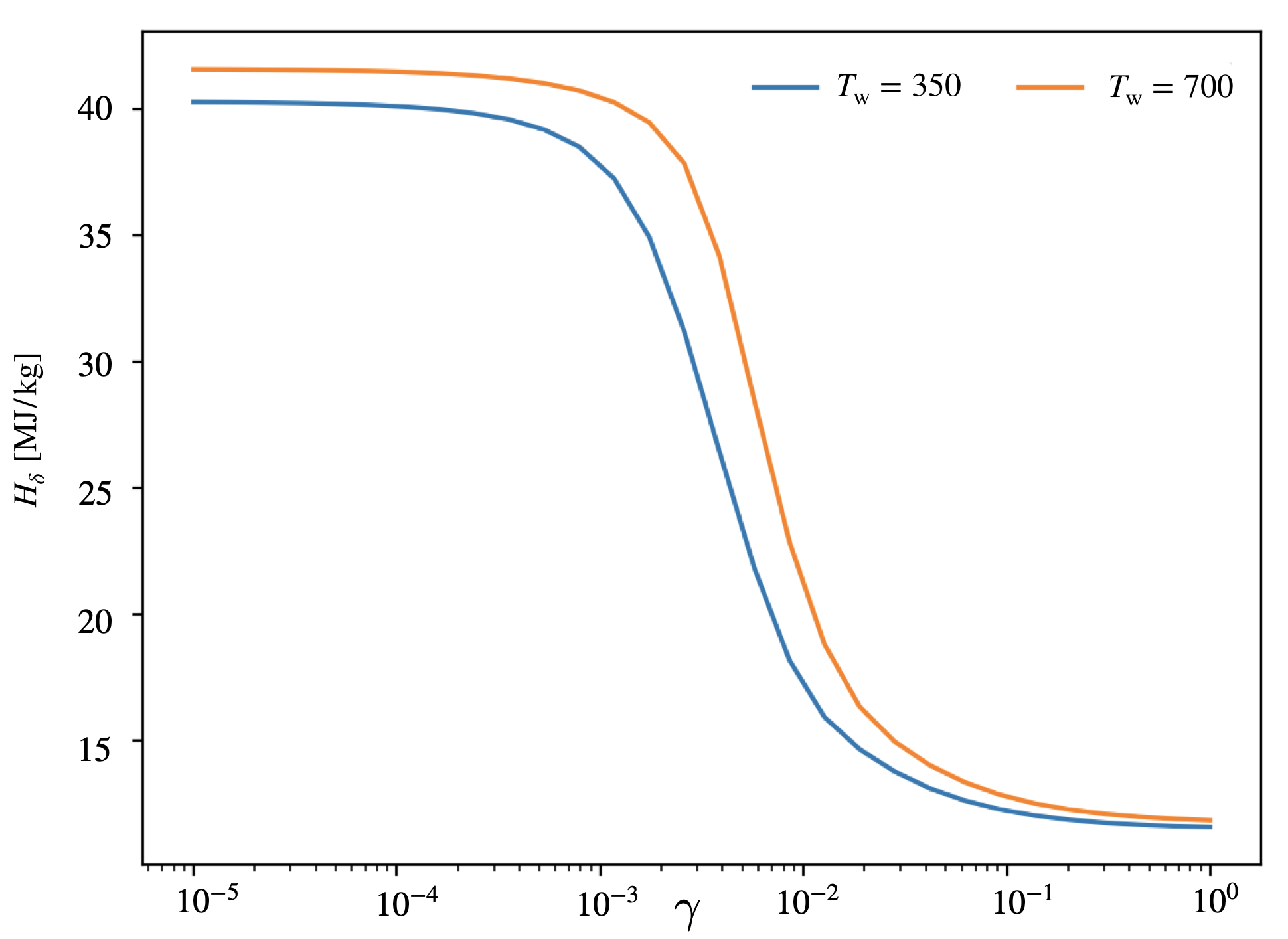} \cr
\end{tabular}}
\caption{Change on the S-shaped curves positions due to changes in heat flux ($q_{\mathrm{w}}$) or wall temperature ($T_{\mathrm{w}}$)}\label{fig:17}
\end{figure}

To understand this better, first we need to take a look back at Sec. 2. In that section, we explain the foundations of the physical phenomena behind this problem. The physical system, represented by the BL code, computes the function $q_{\mathrm{w}}=q_{\mathrm{w}}(H_{\delta}, \Ps, \Pst, \Tw, \gamma)$. For each $\Ps, \Pst, \Tw$ and $q_{\mathrm{w}}$, the system relates the enthalpy $H_{\delta}$ and the catalytic parameter $\gamma$ through an S-shaped curve (see Fig.\ref{fig:17}). During the optimization, the physics allow the S-shaped curves to move when different parameters change, in this case, the resulting heat flux $q_{\mathrm{w}}$ and the wall temperature $T_{\mathrm{w}}$ (Fig.\ref{fig:17}). The pressure quantities play a minor role due to their small uncertainties and the fact that both curves move together when these quantities change, being common for both materials. 

It is also important to take into account that we have information about all the nuisance parameters but the enthalpy $\Hd$ which is not measured, therefore, all the other nuisance parameters try to be close to their measured values as a result of the optimization. The lack of information about $\Hd$ gives more uncertainty in the resulting $H_{\delta}^{\mathrm{opt}}$. In turn, we can think of the optimization algorithm as looking for the optimal $\Hd$ while keeping the other nuisance parameters very close to their measured values (within their prescribed standard deviation). 

\begin{figure}[hbt!]
\centering
\includegraphics[width=0.6\textwidth]{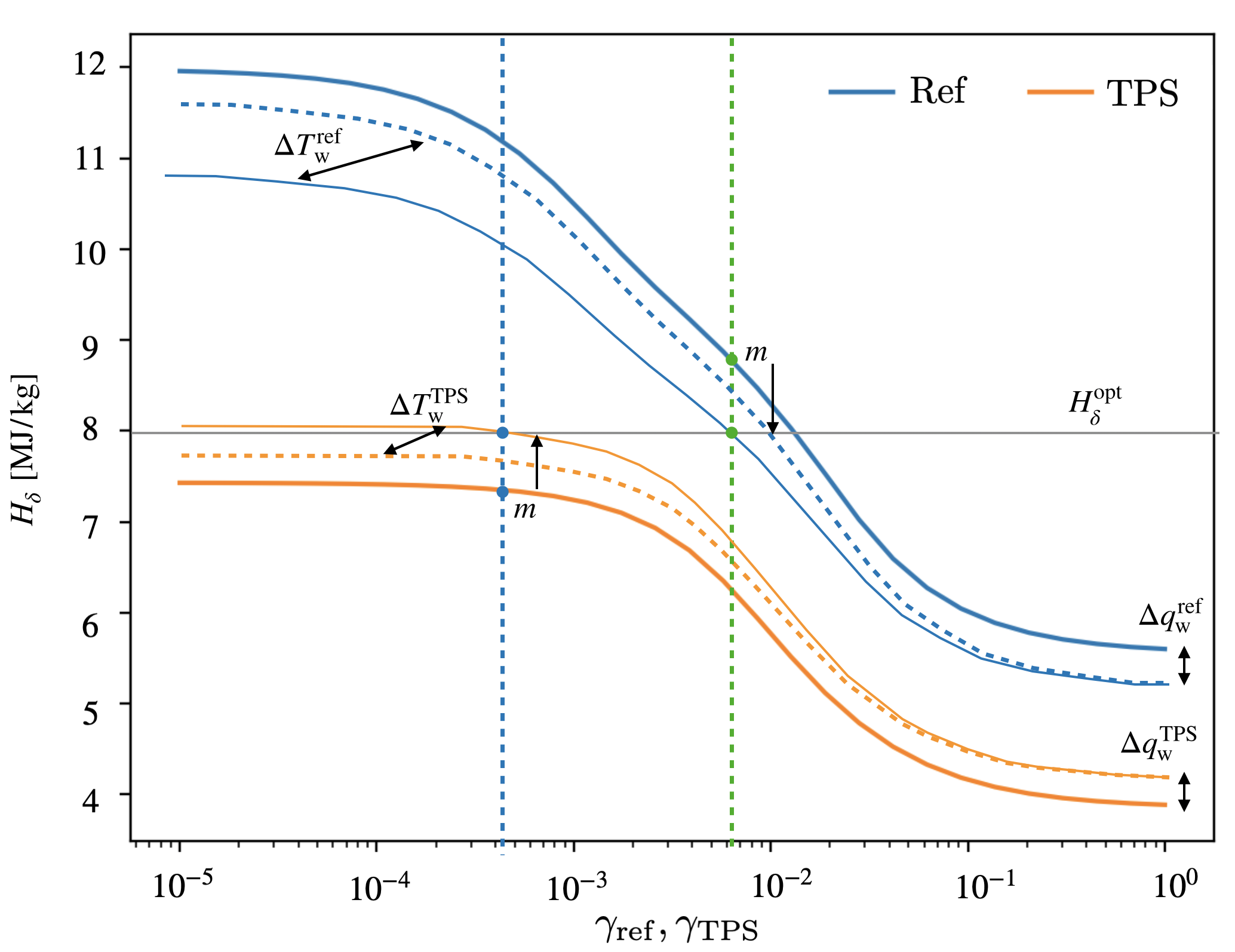}
\caption{Inner workings of the optimization algorithm in terms of the S-shaped curves}
\label{fig:18}
\end{figure}

Fig.~\ref{fig:18} shows the inner workings of the optimization procedure in terms of the physical relations. The thick solid lines represent the S-shaped curves for the reference and TPS material when taking the mean measured values of all the nuisance parameters and the heat flux. The dashed lines represent a change of heat flux from the values of the thick solid lines. The thin solid lines represent the final optimal configuration for the given $\gammar$ and $\gammap$ (vertical dashed lines) for which a change of wall temperature $T_{\mathrm{w}}$ is added to the change of heat flux, transforming the original curves (thick solid lines) to the final optimal curves (thin solid ones). At the pair of $\gammar$ and $\gammap$ where the two thick S-shaped curves have the same $H_{\delta}$, the algorithm finds the model to agree perfectly with the experiments. For the pair of $\gammar$ and $\gammap$ given in Fig.~\ref{fig:18} as an example, the reference and TPS material do not share the same $H_{\delta}$ for their corresponding mean measured values (thick S-shaped curves). The optimization algorithm finds an optimum $H_{\delta}^{\mathrm{opt}}$ which represents a trade-off between the deviations in wall temperatures $T_{\mathrm{w}}$ and heat fluxes $q_{\mathrm{w}}$ with respect to their measured values over their uncertainty range $\sigma$. As the deviation needed to find a common $H_{\delta}$ point for both curves increases, the value of $\mathrm{log} \ \left(L^{\mathrm{opt}}(\bm M \vert \bm \gamma(\bm \xi))\right)$ decreases. In turn, the algorithm performs this for every pair of $\gammar$ and $\gammap$ in our grid, defining the most likely values for the catalytic parameters in light of the experimental data.

\begin{figure}[hbt!]
\centering
\includegraphics[width=0.6\textwidth]{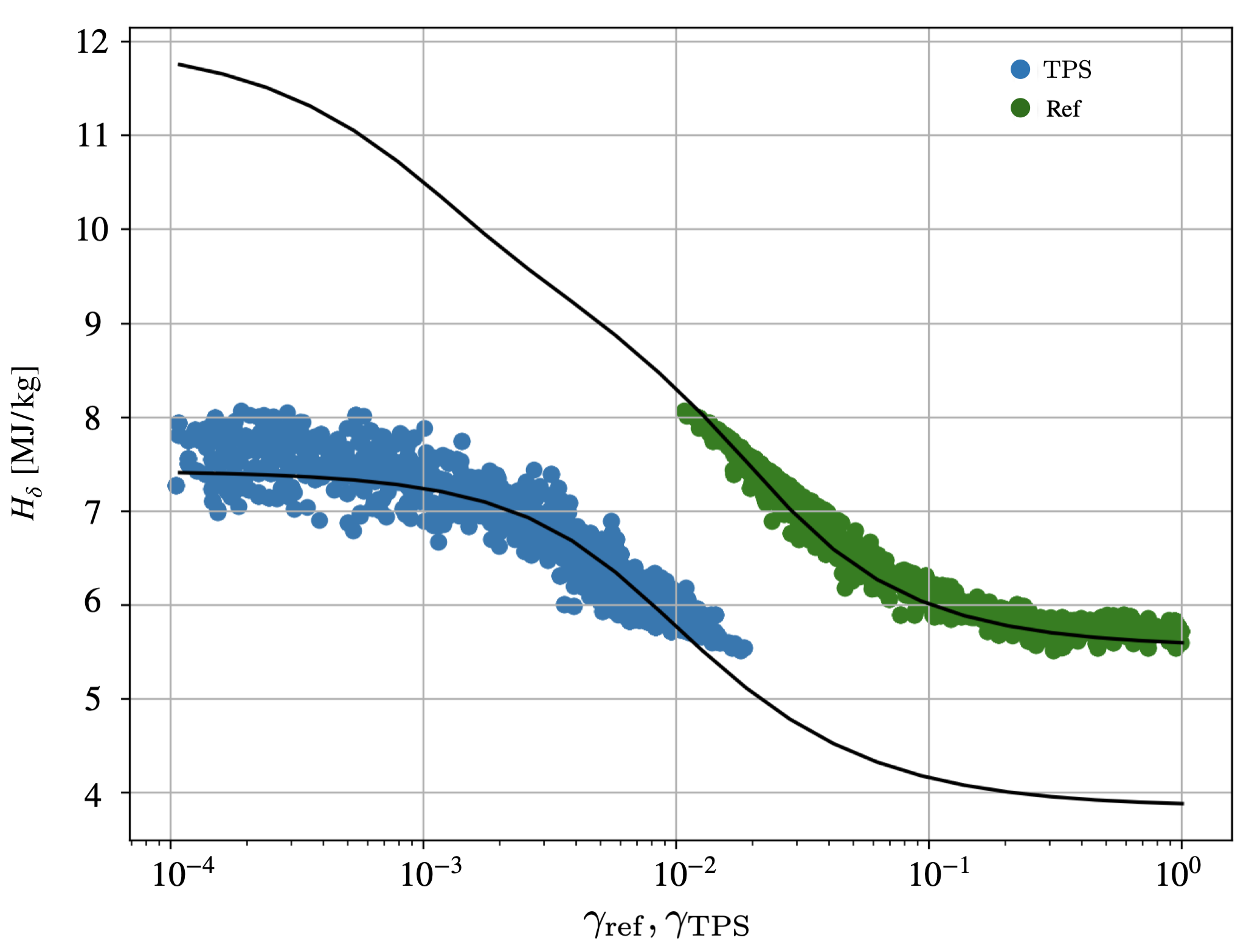}
\centering
  \caption{Posterior samples along the S-shaped curves}
  \label{fig:19}
\end{figure}

As a result, the points sampled by the MCMC algorithm are shown hereafter in Fig. \ref{fig:19}. We can appreciate how the points which maximize the likelihood are the ones falling in the range where both S-shaped curves coincide in enthalpy levels. These points represent the best agreement of the system response with respect to the experimental data. This logic explains the bimodal distribution for the enthalpy and the rest of the optimal parameters.

It is an important exercise to put these results in perspective. We are able to relax some assumptions in our model (catalytic behavior of the reference material) and propose a new functional relationship through the inference framework (optimal likelihood function) but there are other assumptions that remain highly uncertain in within the model. One contributor to such uncertainty is the chemistry of the gas. The chemical state of the gas poses epistemic uncertainties given that different models exist in the literature and are widely adopted. Specifically, the speed of the different reactions considered can play a role in the inference, given that a flow in chemical equilibrium or frozen can produce very different heat fluxes under the same edge conditions. Catalytic activity can also be relegated to be non-influential in the heat flux experienced by the material if the gas chemistry has already taken all the energy contained in the flow and this is likely to occur under higher pressure conditions than the case considered. Nevertheless, the chemistry should be calibrated in dedicated experiments to obtain reliable predictions in the future, as it can impact whether or not the chosen model can explain the experimental data, and this, in turn, influences the calibrated $\gamma_{\mathrm{TPS}}$ obtained. For gas chemistry calibration, dedicated spectroscopic analyses should be included to have conclusive results and be able to learn something about the chemistry of the gas. The experimental data considered in this work would not be enough to make any statement about surface catalysis and chemistry of the gas as it does not provide enough information in itself. The development of this Bayesian framework offers a starting point for future studies for which the experimental data can be thoroughly exploited.

Coupled with the gas chemistry is another uncertain assumption, the thermodynamic state of the gas. Even though this assumption has been validated using spectroscopic measurements \cite{Cipullo} for the test conditions considered, recent numerical investigations \cite{Panesi} suggest that LTE may not hold under different conditions (e.g. lower mass flows). A more extensive use of this framework with dedicated experimental campaigns can help shed light on these issues in the future.

\section{Assessment of experimental methodologies}
\label{Preliminary}

The developed inference methodology can help underpin the important characteristics of testing for catalytic phenomena in TPS, namely, the conditions and/or configurations that can give the most information by decreasing the uncertainty to a minimum. We assess the role of the auxiliary material used for testing, referred until now as ``reference" material. As extending the testing methodology to include three different materials is already a possibility \cite{Viladegut}, we study the information gain with this methodology with synthetic data. In this section, we discuss how choosing different auxiliary materials and performing experiments with more probes impact the outcome of the inference.

\subsection{Influence of the auxiliary testing material}

Apart from the different testing conditions that can be set for a given experiment (power of the facility, static pressure in the testing chamber, mass flow and probe geometry), we also have the freedom of choosing an auxiliary material with which to gather information about the boundary layer edge conditions. 

\begin{figure}[hbt!]
\centering
\includegraphics[width=0.6\textwidth]{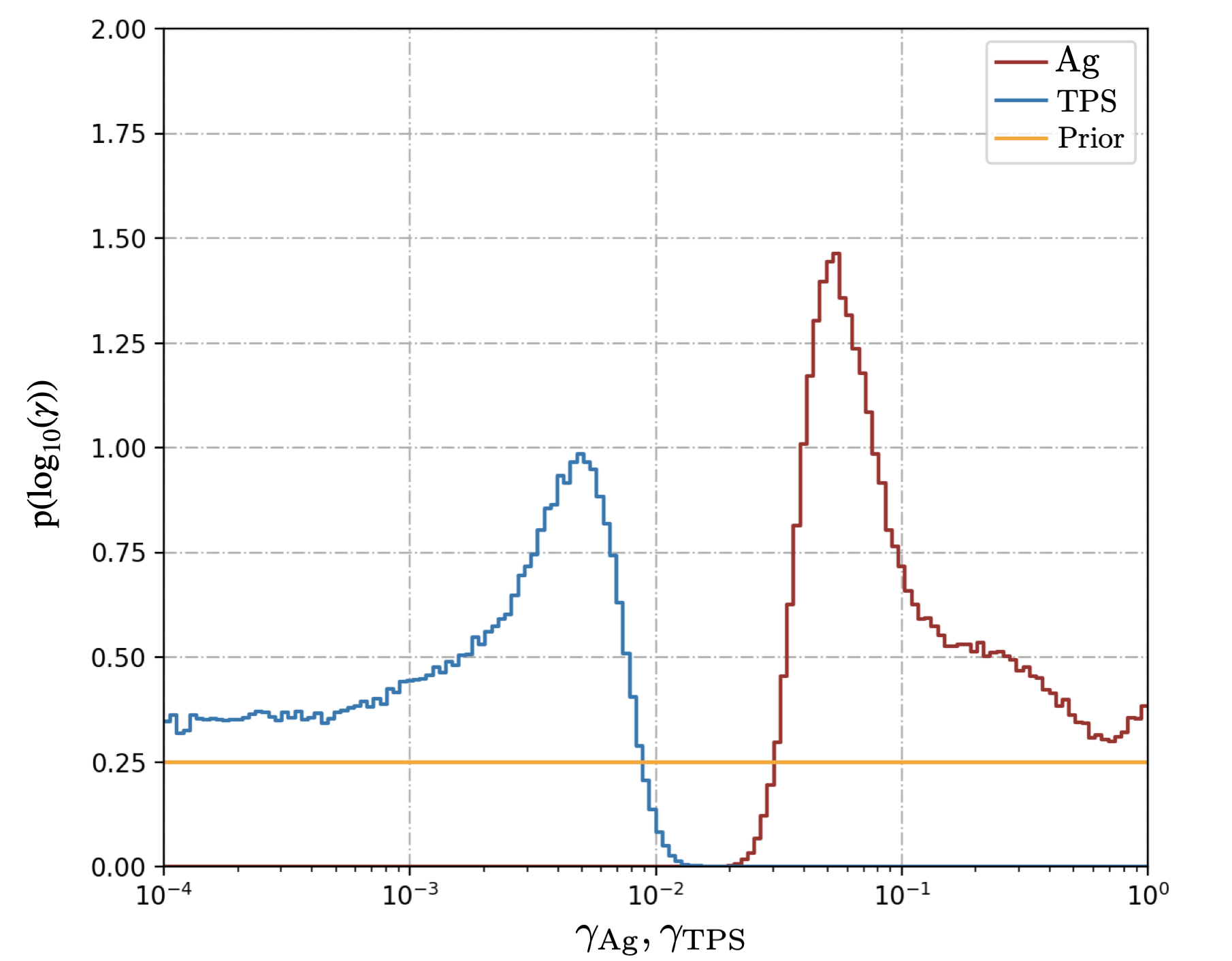}
\centering
  \caption{Marginal posteriors obtained for the TPS material and synthetic silver}
  \label{fig:S1_Ag_marginal}
\end{figure}

As it was explained in previous works \cite{Sanson} and in this work (Sec. \ref{Bayes}), one fundamental uncertainty in the way of rebuilding the TPS catalytic behavior is the fact that the boundary layer edge conditions cannot be estimated accurately if the auxiliary material behavior is not a priori well-known, which is the present case. We explore an assessment of this argument by assuming a high catalytic material (more than the conventional reference of copper) which resembles the catalytic response of a hypothetical probe made with silver~\cite{Viladegut}. We devise synthetic data where the heat flux that the auxiliary material would measure is higher than the previously considered copper, while still keeping the same wall temperature. This way, the only variable that changes from the case of copper to this synthetic case is the catalytic activity at the hypothetical auxiliary material surface. Table \ref{tab:data_syn} shows the data used to simulate this particular case. Results from the inference are depicted in Fig. \ref{fig:S1_Ag_marginal} with the marginal posterior distributions. 

The distributions show the same features than the case study (Fig. \ref{fig:15}) with reduced support and well-defined peaks. We can appreciate that both the supports of the synthetic silver and the TPS are further reduced from the one in Fig. \ref{fig:15} giving a slightly better characterization of these properties.

\begin{table}[hbt!]
\centering
\caption{Synthetic data and uncertainties} 
\label{tab:data_syn}
  \begin{tabular}{c  c  c  c  c  c  c}
    \toprule
    \textbf{Experiment $\mathrm{S_{Ag}}$} & $q_{\mathrm{w}}^{\mathrm{Ag}} \ \mathrm{[kW/m^{2}]}$ & $T_{\mathrm{w}}^{\mathrm{Ag}} \ \mathrm{[K]}$ & $\Ps \ \mathrm{[Pa]}$ & $\Pst \ \mathrm{[Pa]}$ & $T_{\mathrm{w}}^{\mathrm{TPS}} \ \mathrm{[K]}$ & $q_{\mathrm{w}}^{\mathrm{TPS}} \ \mathrm{[kW/m^{2}]}$\\ \midrule
    Mean ($\mu$)& 232 & 350 & 1300 & 75 & 1200 & 91.7\\ 
    Std deviation ($\sigma$)& 7.7 & 11.7 & 1.3 & 1.5 & 40 & 3.05\\ \bottomrule
  \end{tabular}
\end{table}

\begin{figure}[hbt!]
\centering
\includegraphics[width=0.6\textwidth]{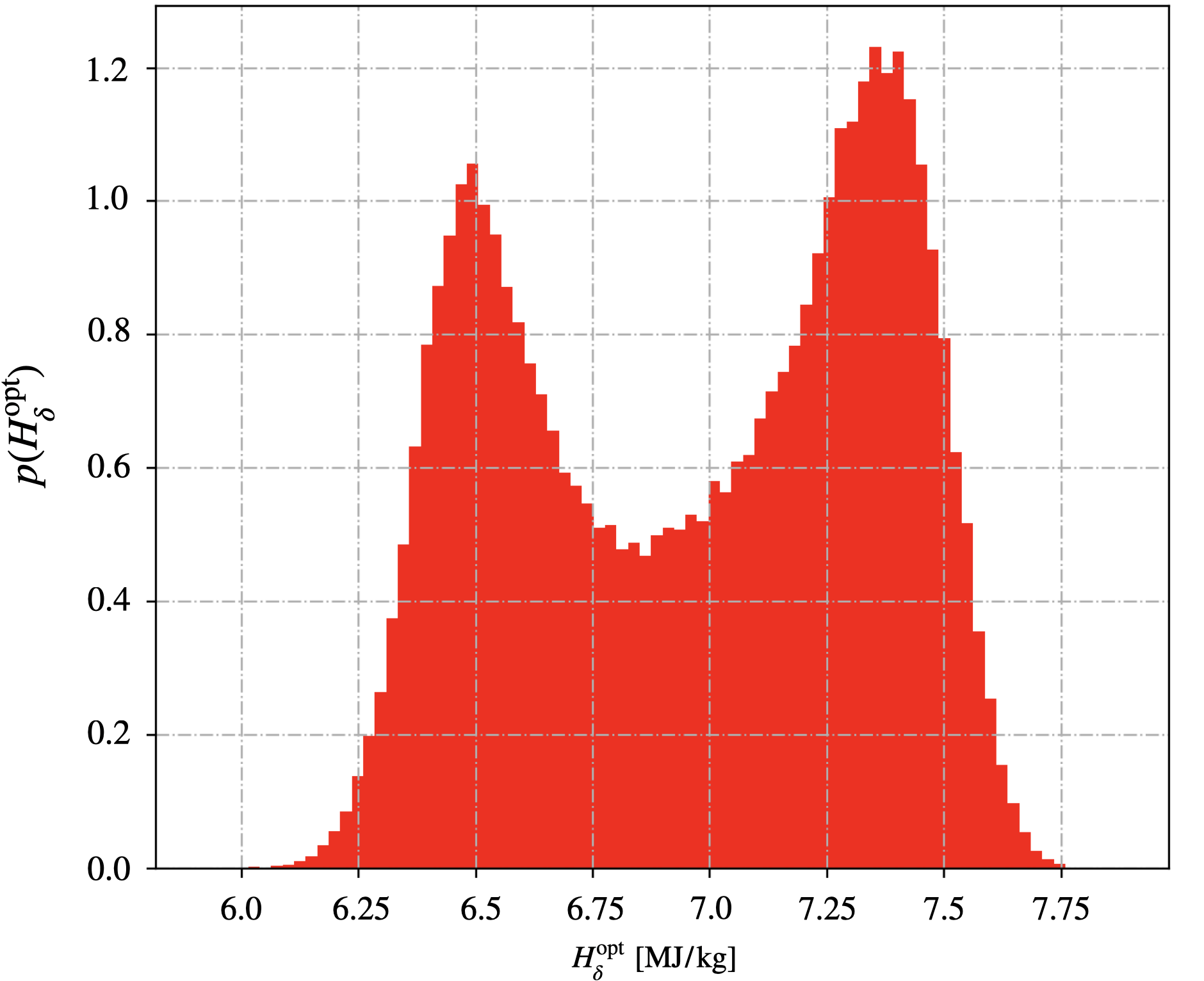}
\centering
  \caption{Optimal enthalpy $H_{\delta}^{\mathrm{opt}}$ distribution obtained by propagating the catalytic parameters posterior of the TPS material and synthetic silver}
  \label{fig:S1_Ag_H}
\end{figure}

To assess the information gain with this particular testing, we need to turn to the enthalpy of the plasma flow and see if we manage to capture this information better with synthetic silver. Figure \ref{fig:S1_Ag_H} shows the distribution for the optimal enthalpy. In this case, it is easy to spot the benefits of changing the auxiliary material to a higher catalytic one. The support is greatly reduced when comparing Fig. \ref{fig:16} and Fig. \ref{fig:S1_Ag_H}. This means that the recovery of the boundary layer edge information is better in the latter case. More information is contained in that experiment than in our case study. Still the characterization of the boundary layer edge conditions could be further improved as it should converge to a unimodal distribution.

\subsection{Extension to a 3-probes testing methodology}

The characterization of catalytic behavior can be further studied with a testing methodology that uses two auxiliary materials instead of one. The information brought by this additional probe is expected to improve the characterization of the boundary layer edge conditions. For the following case study, the three probes seen in this work (ref, TPS and synthetic Ag) are used under the same boundary layer edge conditions to infer their catalytic behaviors and the corresponding enthalpy. This synthetic test case lets us extrapolate the benefits of this methodology to more general cases, prescribing the best practice to reduce the uncertainty on the characterization of catalytic parameters of TPS materials. 

\begin{figure}[hbt!]
\centering
\includegraphics[width=0.6\textwidth]{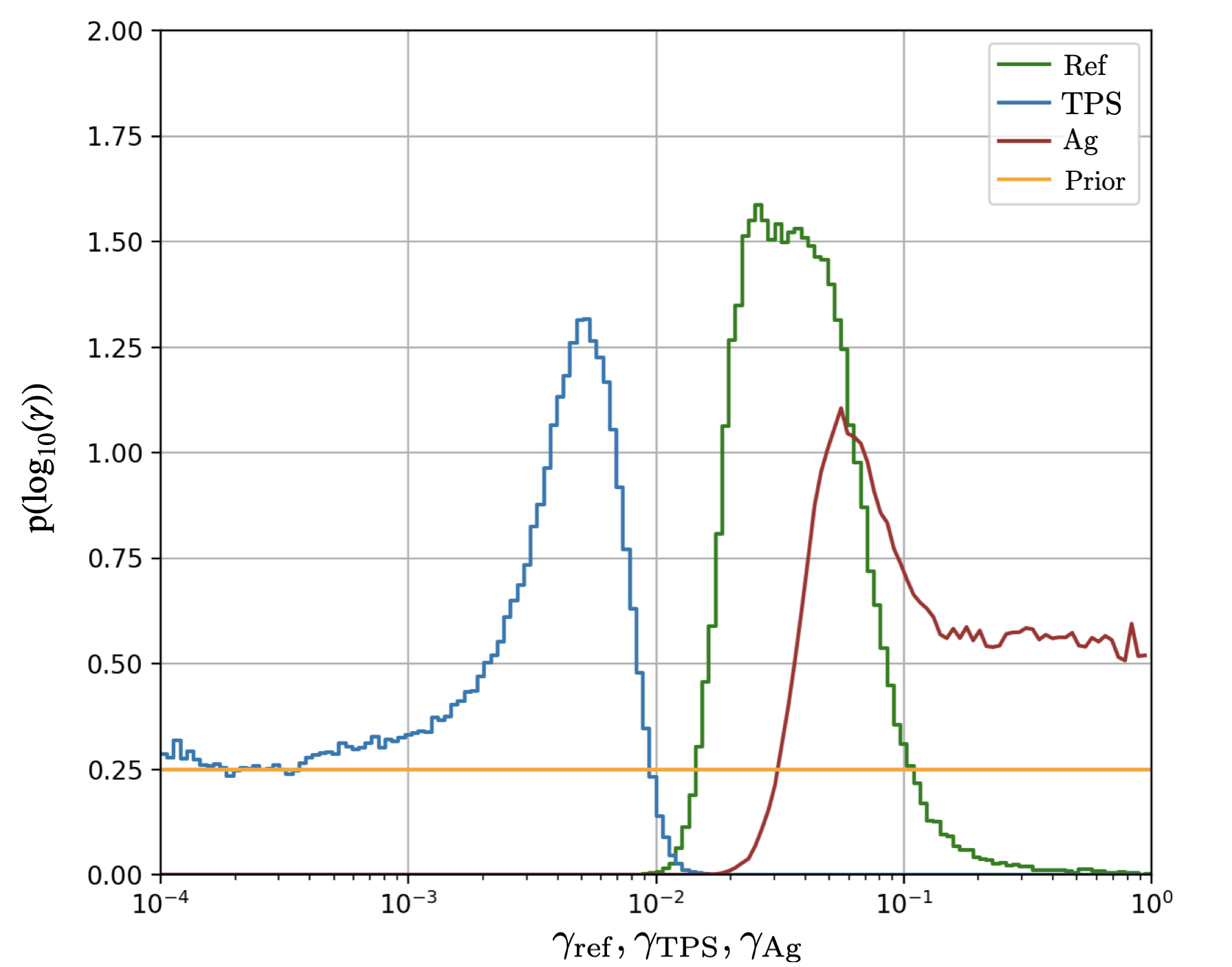}
\centering
  \caption{Marginal posteriors obtained for the TPS material, reference material and synthetic silver}
  \label{fig:S1_3D}
\end{figure}

Fig. \ref{fig:S1_3D} shows the marginal posterior distributions obtained. We can observe that both the TPS and synthetic silver distributions are left almost unchanged from the case where they were tested together (Fig. \ref{fig:S1_Ag_marginal}), although the tail of the synthetic silver distribution shows some growth compared to the previous case. The most notable difference is the posterior distribution of the reference material. The presence of a higher catalytic material increases the information obtained for higher values of the catalytic parameters, reducing, as a consequence, the support for high catalytic values of the reference material. In return, this information gain produces a well-defined peak with a further reduced support. A better characterization of the copper calorimeter is therefore achieved this way.

Table {\ref{tab:table_3D}} depicts the summary statistics of the reference copper and TPS parameters. When compared to Table \ref{tab:table_stats} we can see that the mean value for the reference probe is moved towards lower catalytic values as well as the mean for the TPS. The standard deviation is decreased significantly for the reference probe ($- 80\%$) and for the TPS probe ($- 45\%$) while the MAP values have suffered overall less change.

\begin{figure}[hbt!]
\centering
\includegraphics[width=0.6\textwidth]{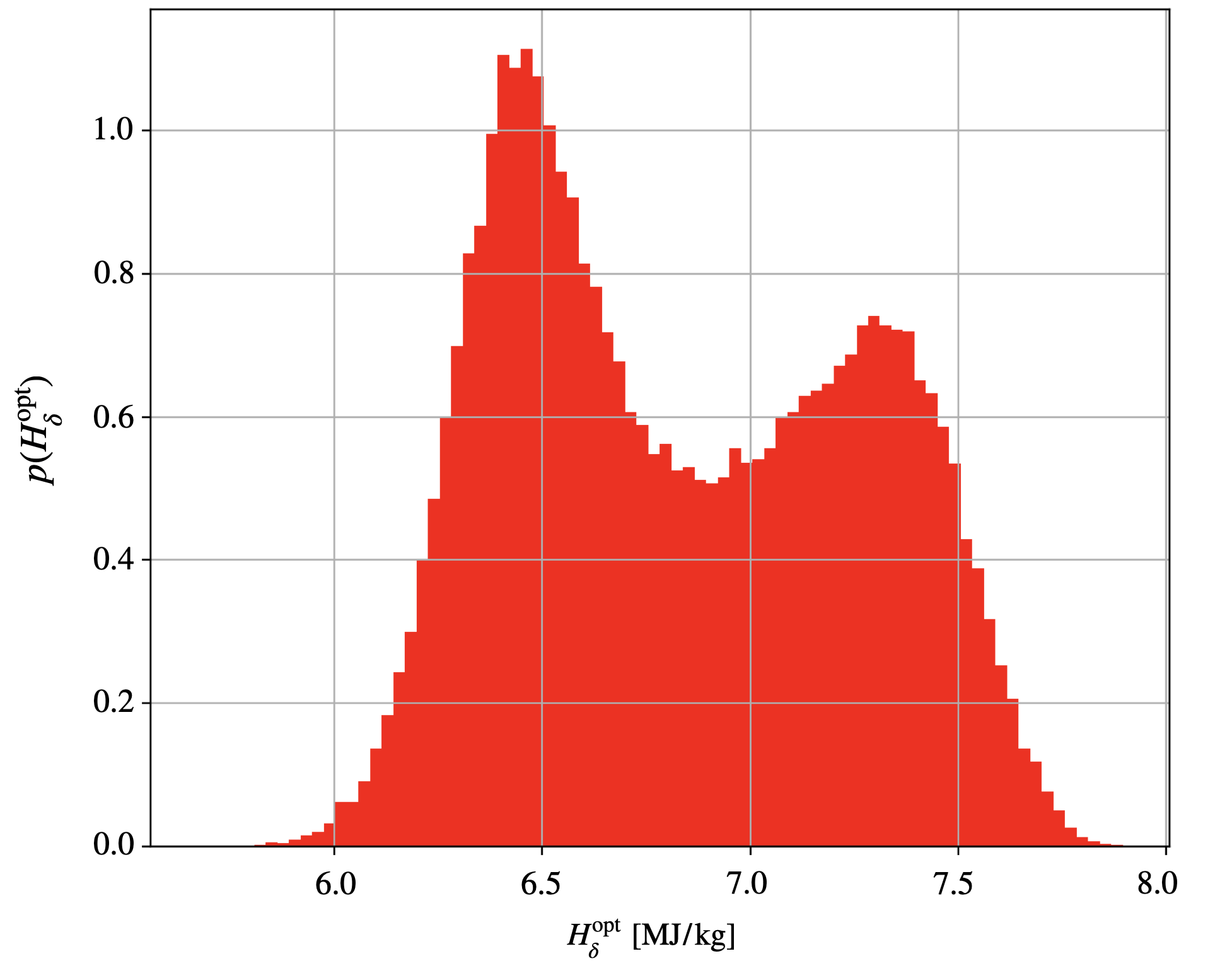}
\centering
  \caption{Optimal enthalpy $H_{\delta}^{\mathrm{opt}}$ distribution obtained by propagating the catalytic parameters posterior of the TPS material, reference material and synthetic silver}
  \label{fig:S1_3D_H}
\end{figure}

\begin{figure}[hbt!]
\centering
\includegraphics[width=0.6\textwidth]{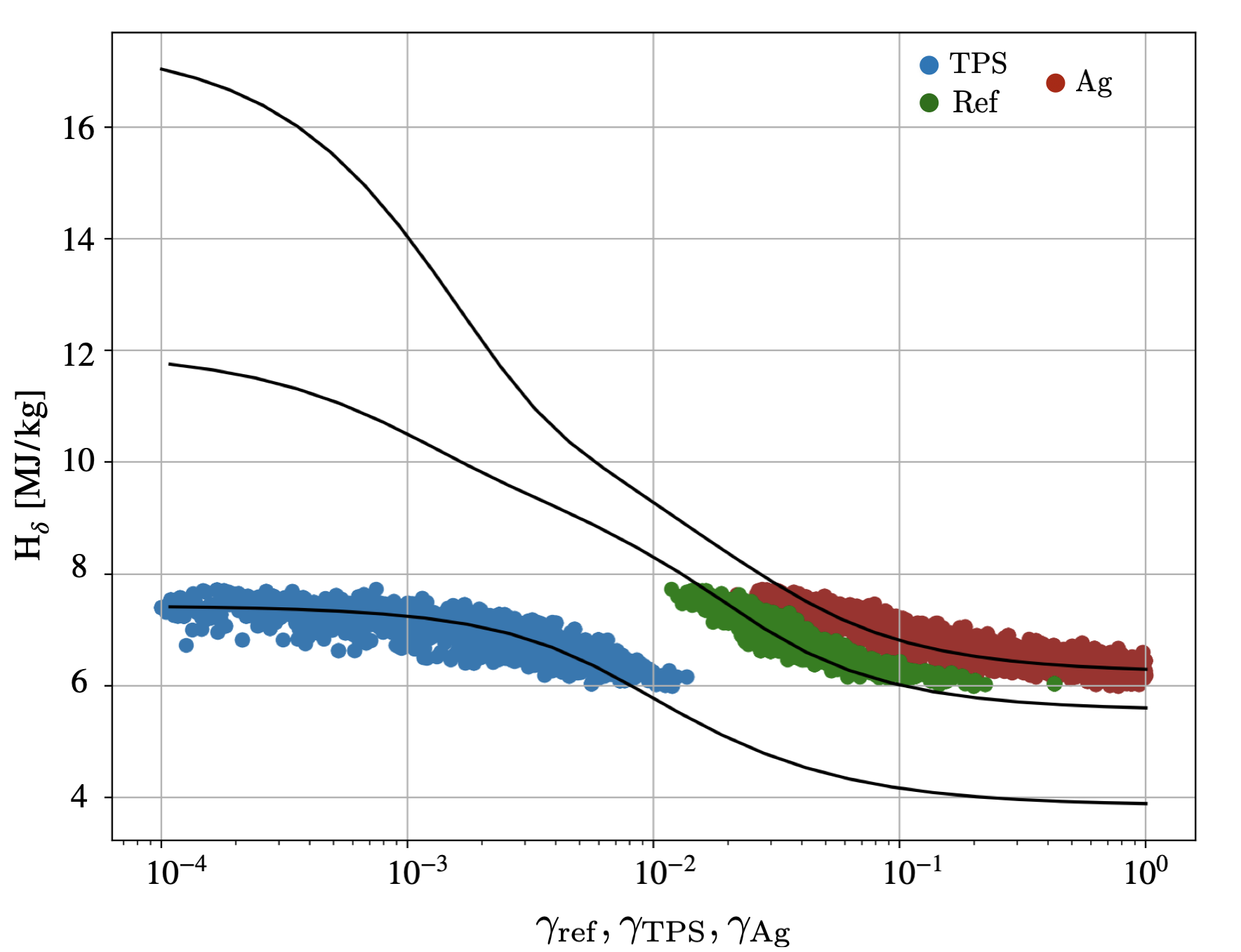}
\centering
  \caption{Posterior samples on the S-shaped curves for the three tested materials}
  \label{fig:S1_3D_curves}
\end{figure}

\begin{table}[hbt!]
\centering
\caption{Posterior statistics for the experiment $\mathrm{S_{Ag}}$ including the reference copper calorimeter} 

  \begin{tabular}{c  c  c  c  c}
    \toprule
    \textbf{Experiment $\mathrm{S_{Ag}}$} & Mean ($\mu$) & Std dev. ($\sigma$) & MAP & CV [$\sigma/\mu$] \\ \midrule
    $\gammar$ & 0.041 & 0.015 & 0.025 & 0.36 \\ 
    $\gammap$ & 0.0027 & 0.0026 & 0.005 & 0.96 \\ \bottomrule

  \end{tabular}
  \label{tab:table_3D}
\end{table}
Turning now to the optimal enthalpy, we can see in Fig. \ref{fig:S1_3D_H} that the resulting support is comparable to the one obtained with just the TPS material and synthetic silver (Fig. \ref{fig:S1_Ag_H}). We can appreciate a slight shift of the modes due to the fact that the error on the measurements of the additional probe weighs in to build the optimized log-likelihood, bringing the optimal enthalpy values closer to the lower plateau where both synthetic silver and copper lay closely together as seen in Fig. \ref{fig:S1_3D_curves}. The amount of information contained in this test case, with three different materials and the previous case with TPS material and synthetic silver is the same and the same support for the optimal enthalpy is retrieved. The materials laying in the extremes of the catalytic spectrum are the ones carrying the information about the boundary layer edge conditions. A closer look at Fig. \ref{fig:S1_3D_curves} when compared to Fig. \ref{fig:19} reveals the fact that the material with a catalytic behavior in between the other two is the best characterized using this methodology. In this regard, the best outcome would be to find a lower catalytic material than the TPS to achieve a better characterization of the material of interest while using copper or silver as the high catalytic auxiliary material.

\section{Concluding remarks and outlook}\label{conclusions}

In this work, we propose a novel Bayesian inference formulation for the calibration of the catalytic parameters of reusable thermal protection materials. The calibration gives estimates of the material catalytic parameter through its posterior probability distribution which can be disseminated for uncertainty propagation analysis. In plasma wind tunnel experiments, the characterization of the reference material behavior plays an important role. In this dedicated framework we disregard the assumption of having a well-characterized reference material, as proven to be in conflict with the respective literature in many cases. The Bayesian approach allows for the simultaneous computation of both materials in the inference process which proves to be more accurate than the conventional sequential approach as shown by Sanson et al. 

Our main contribution is the methodology itself. We derive a likelihood function by considering an optimization problem in the nuisance parameters space, reducing the dimensionality of the likelihood to just the quantities of interest $\gammar, \gammap$. To cope with this computationally demanding likelihood, we propose the use of a surrogate model. GP works quite well for this problem yielding good results with low standard deviations on the chain samples, which represent a good estimation of the error in the surrogate approximation for the posterior samples. In addition, the approach is robust, in the sense that the MCMC sampling method works smoothly for any given conditions. Overall, the optimization formulation presented has the impact of improving considerably the inference results by giving more consistent and accurate posterior distributions of the catalytic parameters when compared with the results of \cite{Sanson}. The main differences being the reduced support and well-defined peaks of the respective posteriors. A more detailed study of the posterior distributions for the case study shows a decrease of $\mathrm{20\%}$ in the standard deviation with respect to the previous work. Subsequently, it is possible to say that the catalytic parameters can be effectively learned from the experimental conditions and under the considered model assumptions. 

The study of different testing methodologies shows that different auxiliary materials have an impact on the information recovered for the free stream enthalpy. This information gain reduces the standard deviation of the catalytic parameters posterior. On these lines, we also study a 3-probes testing methodology and show an overall improvement of the characterization of the reference material up to $\mathrm{80\%}$ with respect to the results of the case study. The 3-probes testing methodology reveals that the ideal possible testing scenario is with three materials where a good characterization is achieved for the one in the middle of the catalytic spectrum. In general, the results achieved by applying this framework help in the discussion about the testing methodologies and the experimental conditions. The most informative testing methodology, combined with the computation of the optimal testing conditions can lead to the proper design of experiments for such thermal protection system. 

In the future, dedicated experimental campaigns, including spectroscopic measurements, can benefit from this work by exploiting the experimental data more thoroughly and adopting the most informative testing methodology. This can help shed light on highly uncertain assumptions as the thermo-chemical state of the gas upon which to improve our model predictions.


\section*{Declaration of competing interests}

The authors declare that they have no known competing financial interests or personal relationships that could have appeared to influence the work reported in this paper.

\section*{Acknowledgements}

This work is fully funded by the European Commission H2020 programme, through the UTOPIAE Marie Curie Innovative Training Network, H2020-MSCA-ITN-2016, Grant Agreement number 722734.

\bibliographystyle{elsarticle-num}
\bibliography{refs}





\end{document}